\documentclass[review]{elsarticle}
\journal{Journal of Power Sources}
\usepackage[utf8]{inputenc} % allow utf-8 input
\usepackage[T1]{fontenc}    % use 8-bit T1 fonts
\usepackage{hyperref}       % hyperlinks
\usepackage{url}            % simple URL typesetting
\usepackage{booktabs}       % professional-quality tables
\usepackage{amsfonts}       % blackboard math symbols
\usepackage{nicefrac}       % compact symbols for 1/2, etc.
\usepackage{microtype}      % microtypography
\usepackage{lipsum}
\usepackage{color,soul}     % for comment hl
\usepackage{amsmath}        % for align % Used for the boldsymbol.

\usepackage{multirow}

\usepackage{setspace}
\usepackage{subfig} % use multiple figures
\usepackage{graphicx}       % use graphics
\usepackage{nomencl}
\renewcommand{\nomname}
\makenomenclature
\usepackage[version=4]{mhchem}

\DeclareMathOperator{\arcsinh}{arcsinh}
\renewcommand{\vec}[1]{\boldsymbol{#1}}
\usepackage{color,soul}
\usepackage{todonotes}

\renewcommand{\textcolor}[2]{#2}
\begin{document}
\begin{frontmatter}
\title{Physics-constrained deep neural network method for estimating parameters in a redox flow battery}

%%%%%%%%%%% AUTHORS %%%%%%%%%%%%%%%
\author[PNNL,UMN]{QiZhi He}
\ead{qzhe@umn.edu}

\author[PNNL]{Panos Stinis}
\ead{panagiotis.stinis@pnnl.gov}

\author[UIUC,PNNL]{Alexandre M. Tartakovsky} 
\ead{amt1998@illinois.edu}

\address[PNNL]{Physical and Computational Sciences Directorate, Pacific Northwest National Laboratory Richland, WA 99354}
\address[UIUC]{Department of Civil and Environmental Engineering, University of Illinois Urbana-Champaign, Urbana, IL 61801}
\address[UMN]{Department of Civil, Environmental, and Geo- Engineering, University of Minnesota, Minneapolis, MN 55455}

%%%%%%%%%%% ABSTRACT %%%%%%%%%%%%%%%
\begin{abstract}
In this paper, we present a physics-constrained deep neural network (PCDNN) method for parameter estimation in the zero-dimensional (0D) model of the vanadium redox flow battery (VRFB). 
% * Approach
In this approach, we use deep neural networks (DNNs) to approximate the model parameters as functions of the operating conditions. This method allows the integration of the VRFB computational models as the physical constraints in the parameter learning process, leading to enhanced accuracy of parameter estimation and cell voltage prediction.
% * Outcome
Using an experimental dataset, we demonstrate that the PCDNN method can estimate model parameters for a range of operating conditions and improve the 0D model prediction of voltage compared to the 0D model prediction with constant operation-condition-independent parameters estimated with traditional inverse methods.
We also demonstrate that the PCDNN approach has an improved generalization ability for estimating parameter values for operating conditions not used in the DNN training.
\end{abstract}


\end{frontmatter}

%\linenumbers

% ===============================================================================
%\input{sections/outline}
%\input{sections/experiment}

%%%%%%%%%%%%%%%%%%%%%%%%%%%%
%%% Introduction %%%
%%%%%%%%%%%%%%%%%%%%%%%%%%%%
\section{Introduction}

%%% ---------------------------------
% Background
% Ref:
% https://ecmiindmath.org/2019/05/18/the-mathematics-of-redox-flow-batteries/
% (Bao et al. 2019)
%(You, 2019)
%%% ---------------------------------
There has been increasing demand for large-scale battery storage systems for renewable power generation, uninterruptible power supplies, emergency backup, and smart grid applications.
Among the various promising rechargeable energy-storage candidates for integration in the grid, a redox flow battery (RFB)
%, which spatially separates the flow-through electrodes, 
is unique owing to its ability to \textcolor{red}{independently determine} the storage capacity and power output, long cycle life, and safety \cite{wang2013recent,noack2015chemistry,weber2011redox}.
One of the most popular and well-studied RFB technologies is the vanadium redox flow battery (VRFB), which was developed in the  1980s \cite{sum1985study,sum1985investigation,skyllas1986new}.
Because the electrolytes used in VRFBs are all vanadium-based (i.e., the electrolyte in the negative electrode contains \ce{V^2+} and \ce{V^3+} ions, whereas that of the positive electrode involves \ce{VO^2+} and \ce{VO2+} ions) instead of being two different electroactive elements, VRFBs offer great reliability during charging and discharging with less cross-contamination.
Because of its unique features, VRFB is closer to commercialization than other RFBs \cite{kear2012development}.
In this work, we use the VRFB technology to demonstrate a proposed machine-learning-based approach for modeling battery performance.

%%% ---------------------------------
% Modeling
%%% ---------------------------------
Many numerical models have been developed as cost-effective approaches to advance the understanding of VRFB technologies \cite{Shah2008}, as well as to improve battery control, optimize operating conditions, and design new materials and architectures for VRFB systems \cite{wang2013recent,Zheng2014}.
The widely used physics-based VRFB models can be divided into zero-dimensional (0D) \cite{Shah2011,Sharma2015,Eapen2019,Lee2019c}, one-dimensional (1D) \cite{Vynnycky2011, Chen2014a}, two-dimensional (2D) \cite{Shah2008,You2009a,Sharma2014}, and three-dimensional (3D) \cite{Ma2011,Xu2013} models based on the considered number of spatial dimensions that concentrations, current density, and electronic/ionic potential vary in.
The 1D, 2D, and 3D models are based on time-dependent partial-differential equations that, in general, must be solved numerically. While more expensive, these models provide a better understanding of the effect of the battery design and operating conditions on battery performance.
On the other hand, the 0D models use lumped (spatially averaged) states (concentrations, current density, and electronic/ionic potential) and are based on the ordinary-differential equations that often allow analytical solutions \cite{Shah2011}. Because of this, the 0D models are much faster and (given a sufficient accuracy) can be used for real-time monitoring and control of the VRFB system \cite{Al-Fetlawi2009,Al-Fetlawi2010}.
%The previous studies show that several lumped VRFB models are capable of predicting the dynamic changes of the cell voltage output under different operating conditions \cite{Shah2011,Sharma2015,Eapen2019,Lee2019c,Vynnycky2011, Chen2014a}.

To be accurate, the 0D model must be properly calibrated. However, the calibration of the 0D models is challenging because (1) some of the ``lumped'' parameters in the 0D model strongly depend on the operating conditions, and (2) these parameters cannot be directly measured in the experiments \cite{You2009a,Ma2012,cheng2020data,Chakrabarti2020,Chen2021}. 
For example, it has been shown that a slower flow rate could lead to less uniform concentrations \cite{Ma2012,Chakrabarti2020} and reduce the values of effective reaction rate coefficients and the effective reactive surface area in 0D models. Also, the study in \cite{You2009a,Bao2019} reported that the increase of applied current causes a higher concentration gradient and polarization in the porous electrodes that affect the lumped parameter values. 
Therefore, the lumped parameters in the 0D model must be adjusted for different operating conditions. 

%%% ---------------------------------
% Estimation
% Ref:
% (Choi, 2020)
% Wang, Biegler, 2018
%%% ---------------------------------
%Most studies on parameter estimation are performed by either additional laboratory experiments or calibration (fitting).
% There are two major approaches in most studies to perform parameter estimation. One is to conduct additional laboratory experiments to direct measure the quantities; another one is based on calibration (fitting). 
%The latter approach is used to estimate, for a specific VRFB model, parameters that cannot be directly measured by experiments.
For a given experiment, the standard inverse procedure to a battery model calibration (regardless of its complexity) is to find a set of parameters that minimize the square difference between the observed and predicted voltages \cite{Shah2008,You2009a,Vynnycky2011,Chen2014a,Wang2018i,Choi2020,Chen2021,Kroner2020,Jokar2016,Zhang2014,bhattacharjee2018precision,chun2020real}.
%strongly depend on the available experimental data
Bayesian methods have also been proposed to compute the posterior parameter distribution given an assumed form of the data likelihood function of the model to match the observed voltage \textcolor{red}{\cite{Wang2018i,kim2019data}}.
%But their performance remain limited when considering the high dimensional parameter space and the nonlinearity of PDE-based models, which lead to prohibitive computational costs \cite{Zhang2014,Choi2020}.
Choi et al. \cite{Choi2020} recently proposed to use a genetic algorithm (GA) for parameter estimation in a semi-2D steady-state VRFB model as an inexpensive alternative to the inverse and Bayesian methods.

However, the existing parameter estimation approaches cannot be used to compute parameters as functions of the operating conditions.  
%
%Thus, the lumped model is critical to capture the spatial variabilities by a averaging fashion. 
%
%%% ---------------------------------
% PCDNN
%%% ---------------------------------
% In this study, we aim to develop a parameter estimation framework for the VRFB system inspired by the idea of physics-informed neural networks (PINNs) \cite{Raissi2019} that have been successfully applied to many inverse modeling and data assimilation problems \cite{Raissi2019,he2020physics,Tartakovsky2020a}.
% In this approach, we propose to use DNNs to learn the model parameters as functions of operating conditions \cite{Goodfellow2016}. In addition, the PINN model outputs a voltage prediction.
% The proposed PCDNN model is trained with experimental data collected under different battery settings and operating conditions.
%\textcolor{red}{
In this study, we aim to develop a machine-learning-enhanced 0D VRFB model with parameters given by the deep neural network (DNN) functions \cite{Goodfellow2016} of the operating conditions.
The key idea of the proposed approach is to train DNNs subject to the VRFB model constraints; we therefore refer to this approach as the physics-constrained DNN (PCDNN) method. The proposed method allows training the DNNs without any measurements of the parameters as functions of the operating conditions. In this work, we only use the measurements of voltage as a function of time during charge-discharge cycles and of the operating conditions to train the DNNs.  
%The proposed PCDNN model is trained with experimental data collected under different battery settings and operating conditions.
%
Once trained, the PCDNN model can predict the parameter values and voltages as a function of time for various operating conditions, including those that were not used for the model training.
We note that the proposed method is different from the physics-informed neural network (PINNs) method \cite{he2020physics,Tartakovsky2020a,tipireddy2019comparative,reyesPRL}
% Raissi2019
for estimating parameters as functions of space and/or state variables in PDE and ODE models.
In the PINN method, both unknown parameters and state variables are models with DNNs and the DNNs are trained jointly subject to the physics constraints in a soft form.
As a result, in the PINN method, the governing equations are satisfied approximately subject to the errors in the training algorithms. Capitalizing on the fact that the 0D model allows an analytical solution, the physics constraints are satisfied exactly in the PCDNN method. We also note that the PCDNN method can be extended to the higher-dimensional (PDE) models of the flow batteries by enforcing the governing PDEs in the PINN method framework. 

\textcolor{red}{We also note that both the PCDNN and PINN methods are different from  ML methods designed to establish a map between input (or, the distribution of inputs) and output parameters controlling and describing the battery performance (e.g., \cite{chun2020real}). One attribute of such models is a relatively high (real or synthetic) data requirement. Here, we train DNNs subject to the physics constraints that significantly reduces the data requirement.} 
%Due to its high approximation capacity, DNNs can to describe highly non-linear functions given by measurement data despite of constant or highly nonlinear relations.
%In this work, we adopt the 0D VRFB model \cite{Eapen2019,Shah2011} to enforce physics in the DNN training. However, the proposed PCDNN framework can be easily extended to other higher-dimensional VRFB models.

%%% ---------------------------------
% Structure
%%% ---------------------------------
This paper is organized as follows: a brief review of the governing equations and associated assumptions in the 0D VRFB model is given in Section \ref{sec:model_0D},
followed by the introduction of the PCDNN approach and the description of the experimental dataset in Section \ref{sec:methods}.
In Section \ref{sec:result}, the PCDNN approach is validated by both synthetic and experimental datasets. %, in comparison to the reference model parameters.
Discussion of the results and conclusions are given in Section \ref{sec:conclusion}.

%%%%%%%%%%%%%%%%%%%%%%%%%%%%
%%% Numerical model %%%
%%%%%%%%%%%%%%%%%%%%%%%%%%%%
\section{Physical VRFB model}\label{sec:model_0D}
\subsection{Model assumptions and equations}\label{sec:concentration}
\begin{figure}[ht!]
	\centering
	\includegraphics[angle=0,width=4in]{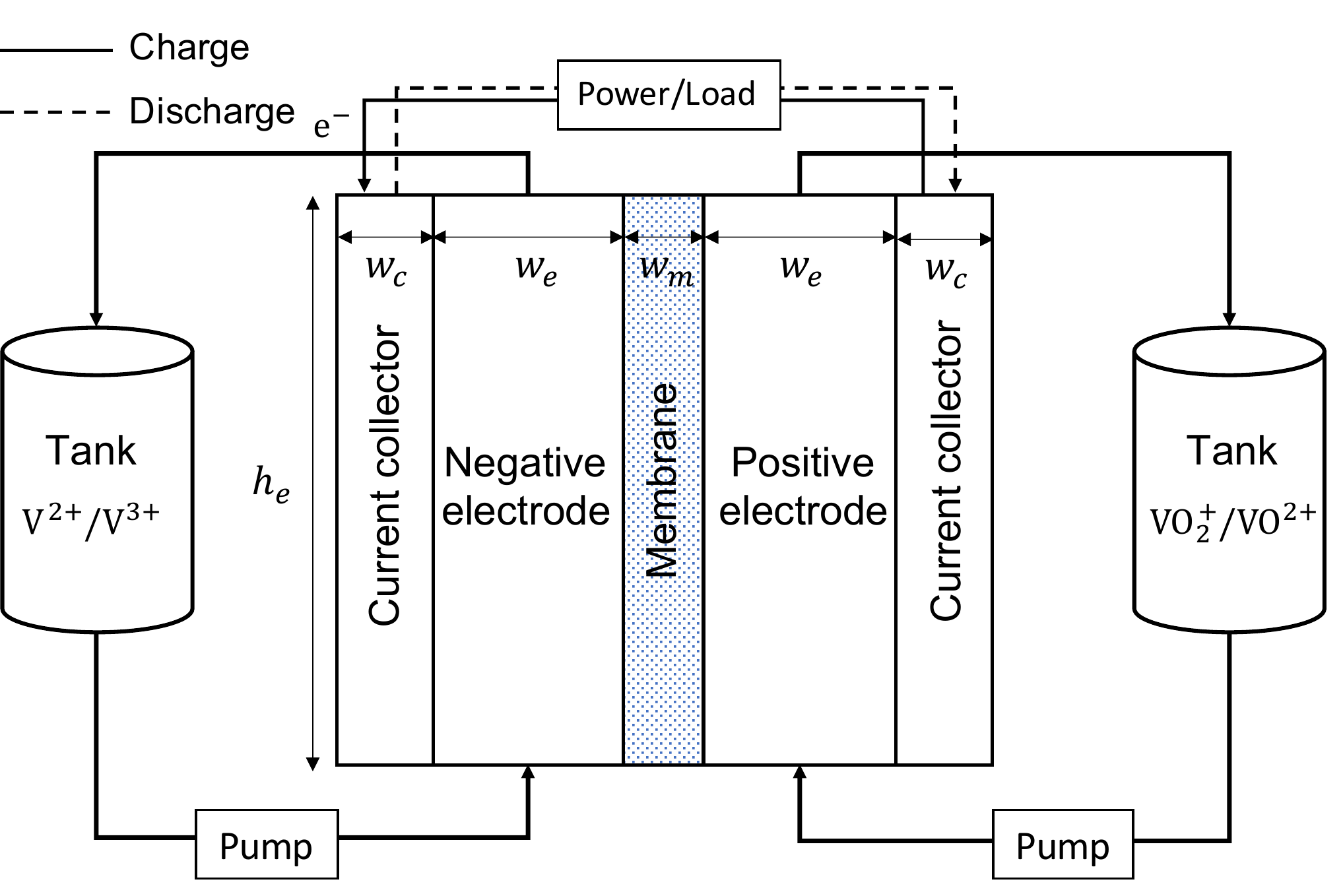}
	\caption{Schematic diagram of a single-cell vanadium redox flow battery (VRFB), which is composed of current collectors, positive and negative electrodes, electrolyte tanks (reservoirs), and an ion exchange membrane.}
	\label{fig:sche_model}
\end{figure}

In this study, the 0D VRFB model \cite{Shah2011} is adopted to describe the reaction kinetics in the electrolytes, electrodes, and membrane of a single-cell VRFB (see Fig. \ref{fig:sche_model}). 
The 0D model is derived based on the mass and energy conservation laws while disregarding the spatial variability of battery properties and electrochemical processes. The latter requires the electrolytes to be sufficiently dilute and be uniformly distributed in the porous electrodes.
In addition, two assumptions are made in deriving the 0D governing equations: (1) the electrolytes are circulated at a constant flow rate between the respective electrodes and reservoirs; and (2) the crossover of vanadium ions through the membrane and any other side reactions are not considered in the model.

%%% ---------------------------------
% Numerical model
%%% ---------------------------------
%\subsection{Physical model}
%
The main chemical reactions in the VRFB can be summarized as 
\begin{subequations}\label{eq:reaction}
	\begin{align}
		& - \text{ve electrode}: \text{V(III)} + \ce{e-} \ce{<=>} \text{V(II)}  \\
		& + \text{ve electrode}:  \ce{VO^2+} + \ce{H2O} \ce{<=>} \ce{VO2+} + 2\ce{H+} + \ce{e-}
	\end{align}
\end{subequations}
Let $c_i, i \in \mathcal{S} = \{\text{V(II)}, \text{V(III)}, \text{V(IV)}, \text{V(V)}, \text{H}^+, \text{H}_2 \text{O}\}$ denote the concentration of active species $i$ in liquid, where  $\text{V(IV)}$ and $\text{V(V)}$ refer to \ce{VO^2+} and \ce{VO2+}, respectively.

The 0D equations governing the evolution of the concentrations of species $c_i$ and their analytical solutions are summarized in  \ref{sec:append-concentration}.
% As shown in \ref{sec:append-concentration}, 
These equations allow the analytical solutions that express the concentration of each species $c_i$ as a function of the state of charge (SOC; defined in \eqref{eq:soc_t} as a function of time $t$): 
\begin{equation}\label{eq:SOC_pinn}
\{c_i (t)\}_{i \in \mathcal{S}} = \{c_i (\text{SOC})\}_{i \in \mathcal{S}} = \left\{
\begin{array}{ll}
% \begin{split}
& c_{\text{V(II)}}  = c_{V}^0 \times \text{SOC} \\
& c_{\text{V(III)}} = c_{V}^0 \times (1-\text{SOC}) \\
& c_{\text{V(IV)}} = c_{V}^0 \times (1-\text{SOC}) \\
& c_{\text{V(V)}}  = c_{V}^0 \times \text{SOC} \\
& c_{\text{H}^+_n}  = c_{\text{H}^+_n}^0 + c_{V}^0 \times \text{SOC} \\
& c_{\text{H}^+_p}  = c_{\text{H}^+_p}^0 + c_{V}^0 \times \text{SOC} \\
& c_{\text{H}_2 \text{O}_p}  = c_{\text{H}_2 \text{O}_p}^0 - (1 + n_d) c_{V}^0 \times \text{SOC}
% \end{split}
\end{array} \right.
\end{equation}
where $c_{V}^0$, $c_{\text{H}^+_n}^0$ ($c_{\text{H}^+_p}^0$), and $c_{\text{H}_2 \text{O}_p}^0$ are the initial concentrations of the vanadium, proton, and water species, respectively. 
The subscripts "$n$" and "$p$" are used to denote the quantities associated with the \textit{negative} and \textit{positive} electrodes,
e.g., $c_{\text{H}^+_p}$ represents the concentration of protons $[\ce{H+}]$ in the positive electrode.
The initial concentration $c_{V}^0$ is equal to the total vanadium concentration in the half cell.
%The derivations of these analytical solutions are given in \ref{sec:append-concentration}.

%%%%%%%%%%%%%%%%%%%%%
%
\subsection{Cell voltage equations}\label{sec:reaction_kinetics}
The cell voltage in VRFB can be computed as  ~\cite{Shah2011,Sharma2015,Eapen2019,Lee2019c}: 
\begin{equation}\label{eq:cell}
E^{cell} = E^{OCV} + \eta^{act} + \eta^{ohm}
\end{equation}
where $E^{OCV}$ is the reversible open circuit voltage (OCV), $\eta^{act}$ is the activation overpotential, and $\eta^{ohm}$ is the ohmic loss. %These components will be further explained in the following.
Eq. (\ref{eq:cell}) ignores the effect of the concentration loss, which is negligible for relatively low current densities. 
%Futhermore, self discharge is neglected. 
%Given an applied current $I$, the cell voltage (or cell potential difference) should be sufficient to drive the forward reaction as well as overcome the irreversible losses. 

In Eq. (\ref{eq:cell}), $E^{OCV}$ can be approximated using a full version of Nernst's equation with the Donnan potential arising across the membrane due to the differences in proton activities between both half-cells \cite{Knehr2011}:
\begin{equation}\label{eq:kin_rev}
E^{OCV} = E_p^0 - E_n^0 
+ \frac{RT}{F} \ln \left( \frac{c_{\text{V(II)}} c_{\text{V(V)}} c_{\text{H}^+_p} c^2_{\text{H}^+_p}}
{c_{\text{V(III)}} c_{\text{V(IV)}}   c_{\text{H}^+_n}  c_{\text{H}_2 \text{O}_p} } \right)
\end{equation}
where $R$, $T$, and $F$ are the ideal gas constant, temperature, and Faraday constant, respectively,
% $c_i  \;  (i \in \mathcal{S} = \{\text{V(II)}, \text{V(III)}, \text{V(IV)}, \text{V(V)}, \text{H}^+, \text{H}_2 \text{O}\})$ are the concentration of species obtained by using the 0D model \cite{Shah2011} introduced in Section \ref{sec:concentration} and \ref{sec:append-concentration}), 
and $E_n^0$ and $E_p^0$ are the standard potentials. Because protons in the positive electrolyte are undergoing a redox reaction to form water, the proton-water redox couple is also considered in Eq. (\ref{eq:kin_rev}).
% Considering the water concentration improves the accuracy of $E^{OCV}$ estimations relative  to the experimental observations \cite{Eapen2019}.

The activation overpotential in Eq. (\ref{eq:cell}) is described by the Butler-Volmer equations~\cite{newman2012electrochemical} as:
% \begin{equation}\label{eq:eta_act}
% \eta^{act} = |\eta_{n}| + |\eta_{p}|
% \end{equation}
\begin{equation}\label{eq:eta_act}
\eta^{act} = \eta_{p} - \eta_{n}
\end{equation}
\begin{equation}\label{eq:eta_neg}
\eta_{n} = -\frac{RT}{\alpha F} \arcsinh \left( \frac{j}{2Fk_n \sqrt{c_{\text{V(II)}} c_{\text{V(III)}} }} \right) 
\end{equation}
\begin{equation}\label{eq:eta_pos}
\eta_{p} = \frac{RT}{\alpha F} \arcsinh \left( \frac{j}{2Fk_p \sqrt{c_{\text{V(IV)}} c_{\text{V(V)}} }} \right)
\end{equation}
where the transfer coefficient $\alpha$ for both electrodes is taken to be 0.5, and $k_n$ and $k_p$ are the rate constants associated with the reactions at the positive and negative electrodes, respectively.
% assumed to be the same t  transfer coefficient is $0.5$ for both electrodes and  for both reduction and oxidation reactions~\cite{Shah2011,Eapen2019}. 
The (local) current density $j$ is calculated by $j=I/A_s$ with $A_s = S V_e$, where $V_e$ is the volume of the electrode and $S$ is the specific surface area.
An empirical expression for the reaction rate coefficients is given in ~\cite{Shah2011}. 
\begin{equation} \label{eq:k1}
k_{n} = k_{n,ref} \exp \left( - \frac{ F E_n^0 (T_{ref})}{R} \left[\frac{1}{T_{ref}} - \frac{1}{T}\right] \right)
\end{equation}
and
\begin{equation} \label{eq:k2}
k_{p} = k_{p,ref} \exp \left( \frac{F E_p^0 (T_{ref})}{R} \left[\frac{1}{T_{ref}} - \frac{1}{T}\right] \right),
\end{equation}
where $k_{n,ref}$ and $k_{p,ref}$ are the reference rate constants for the reactions in Eqs. (\ref{eq:reaction}a) and (\ref{eq:reaction}b) at $T_{ref} = 293$ K, respectively.
% Eqs \eqref{eq:k1} and \eqref{eq:k2} are used to compute the standard reaction rate constants $k_p$ and $k_n$ for given the given reference rate constants.

The ohmic losses associated with the current collector (c), membrane (m), and electrolyte (e) can be expressed as ~\cite{Shah2011,Sharma2015}:
\begin{equation}\label{eq:ohm}
\eta^{ohm} = \left(\frac{2w_c}{\sigma_c} + \frac{w_m}{\sigma_{m}} + \frac{2w_e}{\sigma_e^{eff}}\right) j_{app},
\end{equation}
where $j_{app} $ is the \textit{nominal} current density defined by $j_{app} = I/A_e$, and $w_c$, $w_m$, and $w_e$ are the widths of the collectors, membrane, and electrodes, respectively (as illustrated in Fig. \ref{fig:sche_model}), and $\sigma_c$ is the conductivity of the collector (usually provided by  the manufacturers).
The Bruggeman correction~\cite{Bird2002} is used in the effective conductivity of the porous electrode, given by $\sigma_e^{eff} = \epsilon^{3/2} \sigma_e$.
%We note that the conduction in collectors and electrodes is caused by the movement of electrons, whereas in the membrane it is due to the movement of protons.
Assuming a fully saturated Nafion membrane~\cite{Zawodzinski1993} with $\lambda = 22$, the conductivity of membrane $\sigma_m$ can be expressed as~\cite{Springer1991}:
\begin{equation}\label{eq:sigm_m}
\sigma_m = (0.5139 \lambda - 0.326) \exp \left(1268 \left[ \frac{1}{303} - \frac{1}{T} \right]\right).
\end{equation}

While most of the parameters in the 0D model are independent or weakly dependent on the operating conditions, some parameters, including  $\sigma_e$, $S$, $k_n$, and $k_p$, were found to have a strong dependence on the operating conditions \cite{You2009a,Shah2011,Chen2014a,Eapen2019}.
%However, accurate estimation of these parameters from complex experimental measurements from such varied operating conditions remains challenging. 
In this work, we use the proposed PCDNN method to model these parameters as functions of operating conditions.

\section{Methods \& experiments}\label{sec:methods}
%
%%% ---------------------------------
% PCDNN
%%% ---------------------------------
\subsection{Physics-constrained deep neural networks (PCDNNs)}\label{sec:pinn}
\begin{figure}[ht!]
	\centering
	\includegraphics[angle=0,width=4.5in]{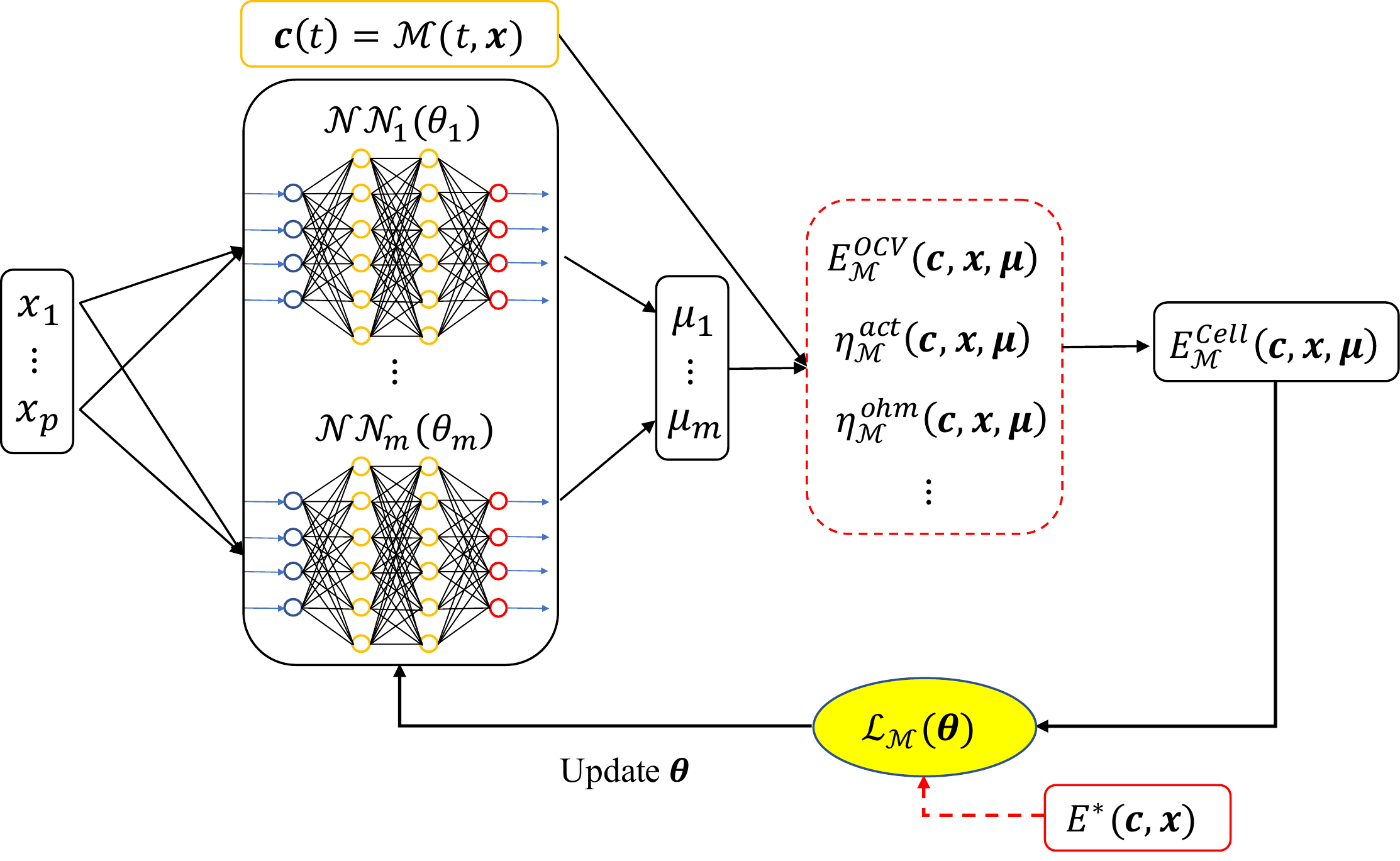}
	\caption{Schematic of the proposed PCDNN approach, where $m$ neural networks are used to relate the operating conditions $\{x_i\}_{i=1}^p$ and the $m$ identifiable parameters $\{\mu_i\}_{i=1}^m$ associated with the physics-based electrochemical model.
		Here, $\mathcal{M}$ denotes a predefined physical model used to compute the concentrations of species $\{c_i \}_{i \in \mathcal{S}}$, open circuit voltage $E_{\mathcal{M}}^{\text{OCV}}$, activation overpotential $\eta_{\mathcal{M}}^{\text{act}}$, and ohmic losses $\eta_{\mathcal{M}}^{\text{ohm}}$. Note that the 0D model described in Section \ref{sec:model_0D} is adopted in this study.
	}
	%{are the output of $m$ neural networks, respective, which associated with the given physics-based electrochemical model (denoted as $\mathcal{M}$) are related with the $p$ operating conditions through 
	%associated with the given   $\{x_i\}_{i=1}^p$ denotes the $p$ operating conditions as inputs to the PINN model, $\mathcal{M}$ indicates a predefined physical model used to compute species concentration, open circuit voltage, overpotentials, and ohmic losses (e.g. 0D model is used in this study).} 
	\label{fig:sche_pinn}
\end{figure}

The general framework of the PCDNN approach is illustrated in Fig. \ref{fig:sche_pinn}, where the model parameters to be estimated are denoted as $ \vec{\mu} = \{\mu_i\}_{i=1}^m$,
and $\vec{x} = \{x_i\}_{i=1}^p$ is the $p$-dimensional vector of the experimental operating conditions for VRFB.
Here, we approximate functional relationships between $\mu_i$ and the operating conditions $\vec{x}$ with fully connected feed-forward DNNs:
\begin{equation}\label{eq:mu_dnn}
	\mu_i (\vec{x}) \approx \hat{\mu}_i (\vec{x}; \theta_i) = \mathcal{NN}_i(\vec{x};\theta_i),  \quad \text{for} \;  i = 1,...,m
	%\mu_i  \in  \{S, k_1, k_2, \sigma_e\}
\end{equation}
where $\mathcal{NN}_i$ is a DNN approximation of $\mu_i$, and $\theta_i$ are the parameters (weights and biases) of $\mathcal{NN}_i$.
We denote $\vec{\theta} = \{\theta_1,...,\theta_m\}$ as the collection of all $\theta_i$.
A brief review of the fully-connected feed-forward DNN architecture is given in  \ref{sec:appendix-DNN-appr}. 

Substituting the DNNs $\hat{\vec{\mu}}(\vec{\theta}) = \{\hat{\mu}_i (\theta_i)\}_{i=1}^m$ in Eq (\ref{eq:cell}) gives the equation for voltage as a function of $\vec{\theta}$, i.e., 
% For any $\vec{x}$, the DNNs $\hat{\vec{\mu}}(\vec{\theta}) = \{\hat{\mu}_i (\theta_i)\}_{i=1}^m$ must satisfy Eq (\ref{eq:cell}), i.e.,
\begin{equation}\label{eq:cell_pinn}
	E^{cell}_{\mathcal{M}}(\vec{c}, \vec{x}, \hat{\vec{\mu}}(\vec{\theta})) = E^{OCV}_{\mathcal{M}}(\vec{c}, \vec{x}, \hat{\vec{\mu}}(\vec{\theta})) + \eta^{act}_{\mathcal{M}}(\vec{c}, \vec{x}, \hat{\vec{\mu}}(\vec{\theta})) + \eta^{ohm}_{\mathcal{M}}(\vec{c}, \vec{x}, \hat{\vec{\mu}}(\vec{\theta}))
\end{equation}
where the subscript $\mathcal{M}$ is used to denote a quantity related to the defined physical models.
% where the subscript $\mathcal{M}$ denotes the 0D electrochemical model for $E^{OCV}$, $\eta^{act}$, and $\eta^{ohm}$ described in Section \ref{sec:reaction_kinetics}.
% Without loss of generality, in this work we use the 0D model to compute these terms, i.e., 
%  $\mathcal{M}$ denotes the 0D model. 
%The model parameters and operating conditions considered in this study are discussed in Sections \ref{sec:conditions} and \ref{sec:parameters}.
%As we can see, for any given $\vec{x}$, the cell voltage outputs $E^{cell}_{\mathcal{M}}(\vec{c}, \vec{x}, \hat{\vec{\mu}})$ exactly satisfy the physical models, thus it is named as a physics-constrained approach.

Considering the concentration--SOC relationship in Eq. (\ref{eq:SOC_pinn}) and the SOC--time relationship in Eq. (\ref{eq:soc_t}), the species concentrations can be written as
% a function of $t$ and  $\vec{x}$, i.e., 
$c_i (t,\vec{x}) = c_i (\text{SOC}(t,\vec{x}),\vec{x})$, and thus, the cell voltage becomes $E^{cell}_{\mathcal{M}}(t, \vec{x}; \vec{\theta})$.

%Once the optimal weight coefficients $\vec{\theta} = \{\theta_1,...,\theta_m\}$  for all DNNs are determined, the learned model parameters $\hat{\vec{\mu}}(\vec{x}; \vec{\theta}^*) = \{\hat{\mu_i} (\vec{x}; \theta_i^*)\}_{i=1}^m $
%%$\hat{\vec{\mu}}(\vec{x}; \vec{\theta}^*) = (\hat{S}(\vec{x}), \hat{k}_n(\vec{x}), \hat{k}_p(\vec{x}), \hat{\sigma}_e(\vec{x}))$
%can be used to predict cell voltage for a given SOC with respect to different operating conditions $\vec{x}$ by using the phyiscs-based 0D electrochemical model (Eq. (\ref{eq:cell})), which can be written as:

%%%%%%%%% Loss function %%%%%%%%%%%%
%%%%%%%%%%%%%%%%%%%%%%%%%%%%%%
The DNN parameters $\vec{\theta}$ in PCDNN are estimated by minimizing the loss function $\mathcal{L}_{\mathcal{M}}$:
%
%The loss function (or objective) $\mathcal{L}_{\mathcal{M}}$ associated with the PINN approach (see Fig. \ref{fig:sche_pinn}) is defined as the following mean square error:
%Using the predicted voltage $E^{cell}_{\mathcal{M}}(\vec{c};\hat{\vec{\mu}}(\vec{x}))$ computed by Eqs. (\ref{eq:cell_pinn}) and (\ref{eq:SOC_pinn}) for a given SOC and operating conditions $\vec{x}=(u, I, c^0_V)$ and considering the mean square error of the voltage, we define a loss (or objective) function with the measurements $E^*(t,\vec{x})$ as follows:
%
\begin{equation}\label{eq:loss_pinn}
\vec{\theta} = \text{arg}\min_{\vec{\theta}^*}	\mathcal{L}_{\mathcal{M}} (\vec{\theta}^*) = \frac{1}{N^{x} N^t_q} \sum_{q=1}^{N^{x}} \sum_{l=1}^{N^t_q} [E^{cell}_{\mathcal{M}}(t_l, \vec{x}_q; \vec{\theta}^*) - E^*(t_l,\vec{x}_q)]^2
\end{equation}
where $E^*$ are the experimental measurements of the cell voltage,
$N^x$ is the number of experiments with different operating conditions,
$N^t_q$ is the number of measurements within a charge-discharge process in each experiment,
% in the SOC-V (or time-V) curves associated with the operating condition $\vec{x}_q$,
and $t_l$ ($l=1,...,N_q^t$) are times where the measurements are collected.
%he time $t$ and SOC is related by using Eq. (\ref{eq:soc_t}), thus they are interchangeable in this study). 
If we define a new array $\vec{z} = (t, \vec{x})$ to encode both the input time variable $t$ and the operating conditions $\vec{x}$, the loss function can be simplified as:
\begin{equation}\label{eq:loss_pinn2}
	\mathcal{L}_{\mathcal{M}} (\vec{\theta}) = \frac{1}{N} \sum_{n=1}^{N}  [E^{cell}_{\mathcal{M}}(\vec{z}_n,\vec{\theta}) - E^*(\vec{z}_n)]^2,
\end{equation}
where $N$ is the total number of the measurements.
We use gradient descent minimization algorithms, including L-BFGS-B \cite{Byrd1995} and Adam \cite{Kingma2015} methods, to minimize $\mathcal{L}_{\mathcal{M}}$.
To alleviate potential overfitting, we adopt the $L_2$ regularization on $\vec{\theta}$ \cite{Goodfellow2016} with a penalty parameter $10^{-8}$ in the loss function.
Once $\vec{\theta}$ is found, the model parameters can be computed using the DNN $\hat{\vec{\mu}}(\vec{x};\vec{\theta})$ for any given operating conditions $\vec{x}$.  
%

%% New for operating conditions..

In this work, we attempt to learn four model parameters including the specific area $S$, the reaction rate constants $k_n$ and $k_p$, and $\sigma_e$, such that $\vec{\mu} = (S, k_n, k_p, \sigma_e)$. 
Furthermore, we assume that the parameters $\vec{\mu}$ depend on the following operating conditions: the average electrolyte flow velocity $\tilde{u}$ ($\tilde{u} = \omega / A_{in}$, see \ref{sec:append-concentration}); the applied current $I$; and the initial vanadium concentration $c^0_V$. Therefore, the vector of operating conditions is $\vec{x}=\{\tilde{u}, I, c^0_V\}$. 
% , see \ref{sec:append-concentration}
%The equations of the concentrations related to these three inputs are detailed in  \ref{sec:append-concentration}.
% We note that while $\vec{\mu}$ only depends on $\vec{x}$, the predicted cell voltage in Eq. (\ref{eq:cell_pinn}) depends on operating conditions in Table \ref{table:PNNL_exp} by means of the 0D  model $\mathcal{M}$.
The choice of parameters and operating conditions is based on the reported dependence of  $S$, $k_n$, $k_p$, and $\sigma_e$ on $\tilde{u}$, $I$, and  $c^0_V$, as discussed in Section \ref{sec:reaction_kinetics}.  
Furthermore, it is shown in \cite{Choi2020} that the voltage prediction has a relatively high sensitivity with respect to the selected parameters.

%%
% \textcolor{red}{"
% In contrast to purely data-driven approaches, the PCDNN prediction $E^{cell}_{\mathcal{M}}$ satisfies the physics-based model $\mathcal{M}$.
% Moreover, with the help of automatic differentiation \cite{Baydin2015,Raissi2019,he2020physics}, the proposed PCDNN approach can be easily extended to high-dimensional, multiphysics electrochemical models $\mathcal{M}$ involving partial differential equations \cite{Shah2008}.
% "}

%%% ---------------------------------
%  Numerical details
%%% ---------------------------------
\subsection{Parameter normalization}\label{sec:numerical_detail}
%Nevertheless, multiple parameter estimation remains difficult when using gradient descent because the significant difference in magnitude between the parameters to be estimated could lead to a ill-conditioned optimization problem.
Because the parameters $\vec{\mu}$ are positive-valued functions, we introduce the following expression for $\hat{\mu}_i$ in \eqref{eq:mu_dnn} to enforce its positivity in the PCDNN model:
\begin{equation}\label{eq:mu_dnn_scale}
	\hat{\mu}_i (\vec{x}) = \mu_i^0 \exp\left ( {y_i (\vec{x})} \right), \quad \text{for} \;  i = 1,...,4,
%	\mathcal{NN}_i(\vec{x};\theta_i),  
\end{equation}
where the variables $y_i(\vec{x})$ are approximated with DNNs as
\begin{equation}\label{eq:mu_dnn_scale2}
	y_i (\vec{x}) = \hat{y}_i (\vec{x},\theta_i), \quad \text{for} \;  i = 1,...,4.
%	\mathcal{NN}_i(\vec{x};\theta_i),  
\end{equation}
and $(\mu_1^0, \mu_2^0, \mu_3^0, \mu_4^0) = (S^0, k_n^0, k_p^0, \sigma_e^0)$ are the predefined \textit{baseline} parameters that we define as the parameters' values taken from the literature (or, this could be the averages of the parameter values that are reported in the literature).

\subsection{Experiment}\label{sec:data_pnnl}
%\textcolor{red}{More experimental descriptions?} \\
\begin{table}[htb]
	\centering
%    	\small{
		\caption{Summary of operating conditions and VRFB parameters for the 12 experiment cases (ID 1-12) for a single cell structure conducted at Pacific Northwest National Laboratory (PNNL). The membrane width is $w_m = 1.27 \times 10^{-2}$ for Nafion 115 and $w_m = 5.08 \times 10^{-3}$ cm for Nafion 212, respectively. The operating temperature is assumed to be $T = 298K$.}
		\label{table:PNNL_exp}
		\scalebox{0.7}{
			\begin{tabular}{cccccccccc}
				\toprule
				Exp. Case  &  $c^0_{V}$  & $c_{\text{H}^+_p}^0$  &  	$ c_{\text{H}^+_n}^0$ &  $ c_{\text{H}_2 \text{O},p}$  & $ c_{\text{H}_2 \text{O},n}$ & $ \omega$  &  I  & $V_r$  &  Membrane \\
				 ID			  & [mol $\text{m}^{-3}$]  & [mol $\text{m}^{-3}$]  & [mol $\text{m}^{-3}$]  &  [mol $\text{m}^{-3}$]  &  [mol $\text{m}^{-3}$]    &  [$\text{ml}$ $ \text{min}^{-1}$]  & [$ \text{A}$ ]   & [$\text{m}^{3}$] &  \\
				\hline
				1 & $1.5 \times 10^{3}$  & $3.85 \times 10^{3}$  & $3.03 \times 10^{3}$ & $4.46 \times 10^{4}$ & $4.61 \times 10^{4}$ & 30 & 0.5   & $2 \times 10^{-5}$ &  Nafion 115\\ % 1_5VOSO4-3-5H2SO4-N115
				2 & $1.5 \times 10^{3}$  & $3.85 \times 10^{3}$  & $3.03 \times 10^{3}$ & $4.46 \times 10^{4}$ & $4.61 \times 10^{4}$ & 20 & 0.75   & $8 \times 10^{-5}$ &  Nafion 115\\ % 1_5VOSO4-3_5H2SO4-1
				3 & $2 \times 10^{3}$  & $5 \times 10^{3}$  & $3 \times 10^{3}$ & $4.75 \times 10^{4}$ & $4.95 \times 10^{4}$ & 20 & 0.5   & $5 \times 10^{-5}$ &  Nafion 115\\ % Bin-2-2-5V-H2O2-1h-N115-04232015
				4 & $2 \times 10^{3}$  & $5 \times 10^{3}$  & $3 \times 10^{3}$ & $4.75 \times 10^{4}$ & $4.95 \times 10^{4}$ & 20 & 0.69  & $4.5 \times 10^{-5}$ &  Nafion 115\\ % Bin-2-2-5V-N115-0_05Nb-01222013
				5 & $2 \times 10^{3}$  & $5 \times 10^{3}$  & $3 \times 10^{3}$ & $4.75 \times 10^{4}$ & $4.95 \times 10^{4}$ & 20 & 0.75  & $4.5 \times 10^{-5}$ &  Nafion 115\\ % Bin-2-2-5V-N115-0_05NbW-02152013-3 (new-ID: 7)
				6 & $2 \times 10^{3}$  & $5 \times 10^{3}$  & $3 \times 10^{3}$ & $4.75 \times 10^{4}$ & $4.95 \times 10^{4}$ & 20 & 1.5   & $4.5 \times 10^{-5}$ &  Nafion 115\\ % Bin-2-2-5V-N115-air-heated-01162012 (new-ID: 9)
				7 & $2 \times 10^{3}$  & $5 \times 10^{3}$  & $3 \times 10^{3}$ & $4.75 \times 10^{4}$ & $4.95 \times 10^{4}$ & 20 & 0.5   & $5 \times 10^{-5}$ &  Nafion 212\\ % Bin-2-2-5V-N212-03032013 (new-ID: 11)
				8 & $2 \times 10^{3}$  & $5 \times 10^{3}$  & $3 \times 10^{3}$ & $4.75 \times 10^{4}$ & $4.95 \times 10^{4}$ & 20 & 0.4   & $3 \times 10^{-5}$ &  Nafion 212\\ % JK-N212-V3-V4-225-cycle (new-ID: 13)
				9 & $2 \times 10^{3}$  & $5 \times 10^{3}$  & $3 \times 10^{3}$ & $4.75 \times 10^{4}$ & $4.95 \times 10^{4}$ & 20 & 0.4   & $2.5 \times 10^{-5}$ &  Nafion 212\\ % JK_N212_10CentNH4Cl_Cycle (new-ID: 14)
				10 & $2 \times 10^{3}$  & $5 \times 10^{3}$  & $3 \times 10^{3}$ & $4.75 \times 10^{4}$ & $4.95 \times 10^{4}$ & 20 & 0.5   & $4\times 10^{-5}$ &  Nafion 212\\ % JK_N212_AceticAcid (new-ID: 15)
				11 & $2 \times 10^{3}$  & $5 \times 10^{3}$  & $3 \times 10^{3}$ & $4.75 \times 10^{4}$ & $4.95 \times 10^{4}$ & 20 & 1.0   & $2\times 10^{-5}$ &  Nafion 212\\ % JK_N212_SGLThick_10cm2_Mix (new-ID: 17)
				12 & $1.5 \times 10^{3}$  & $3.85 \times 10^{3}$  & $3.03 \times 10^{3}$ & $4.46 \times 10^{4}$ & $4.61 \times 10^{4}$ & 20 & 0.4   & $3\times 10^{-5}$ &  Nafion 212\\ % JK_N212_V4+Sulfate_V3+Sulfate_Blank_Cycle011917 (new-ID: 18)
			\bottomrule
			\end{tabular}
%		}
		}
\end{table}

In this study, we focus on analyzing 12  VRFB experiments \cite{Bao2019} that use a cell like the one depicted in Fig. \ref{fig:sche_model}. The experimental settings and conditions of each case are summarized in Table \ref{table:PNNL_exp}.
In these experiments, the electrolyte was made of 1.5M $\text{VOSO}_4$ (Aldrich, 99\%) dissolved in 3.5M $\text{H}_2 \text{S} \text{O}_4$ solution (Aldrich, 96-98\%).
%were used to create  for the VRFB cell.
All experiments were run at room temperature.
The VRFB single cell was composed of two current collectors with the thickness of $w_c = 1.5$ cm, two carbon felt electrodes (each $5 \times 4 \times 0.4$ cm), two reservoir tanks, and a membrane. The electrode area is $A_e = 20$ $\text{cm}^2$.
The Nafion 115 and 212 membranes with thickneses $w_m = 1.27 \times 10^{-2}$ and  $5.08 \times 10^{-3}$ cm, respectively, were used.
% The Other parameters are listed in Table \ref{table:PNNL_exp}.
%, the distinct settings of operating conditions and reservoir volume ($V_r$) used for the VRFB system in these 12 experiments are summarized in Table \ref{table:PNNL_exp}.

\begin{figure}[htb!]
	\centering
	\subfloat[] {\includegraphics[width=2.5in]{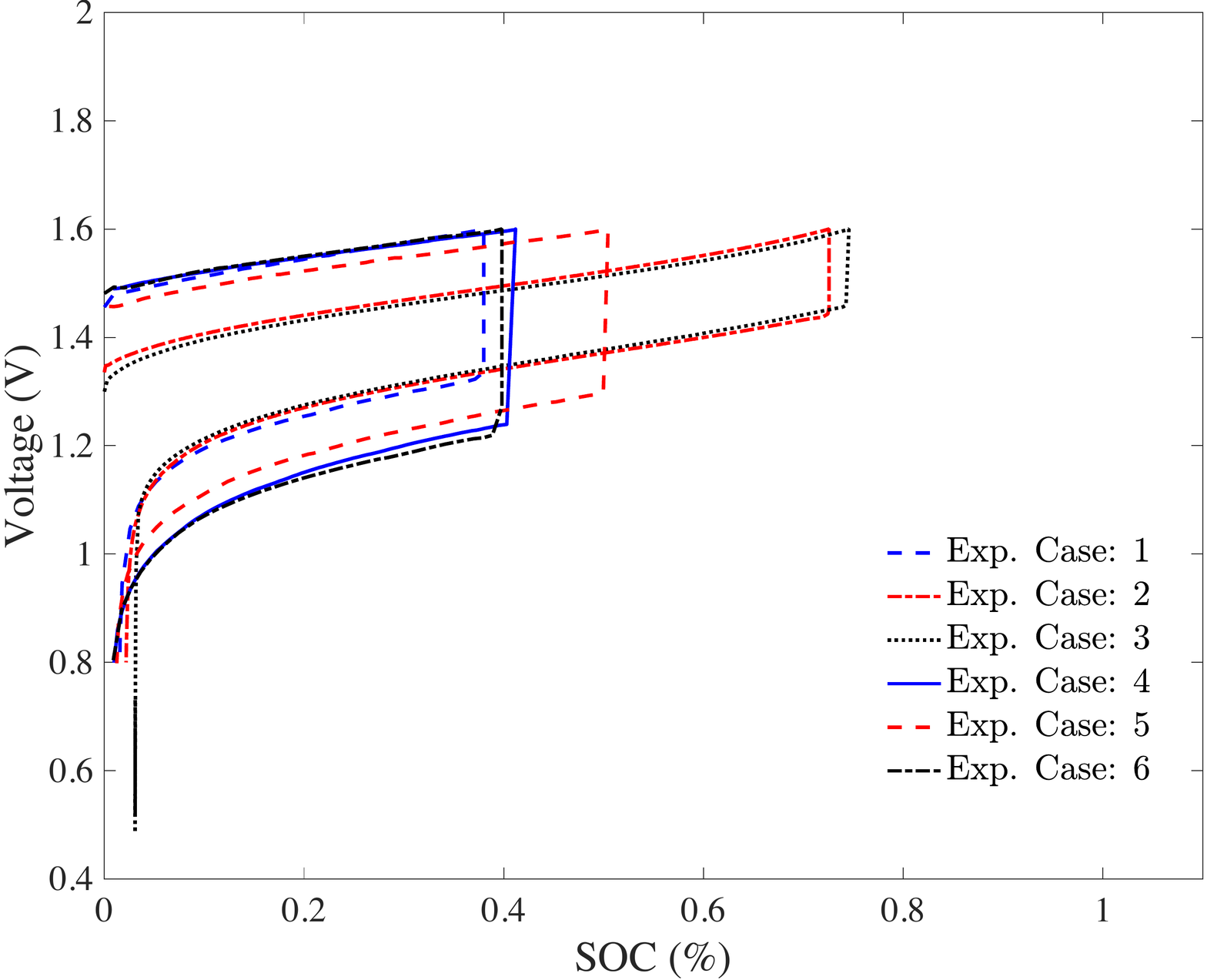}}
	\subfloat[] {\includegraphics[width=2.5in]{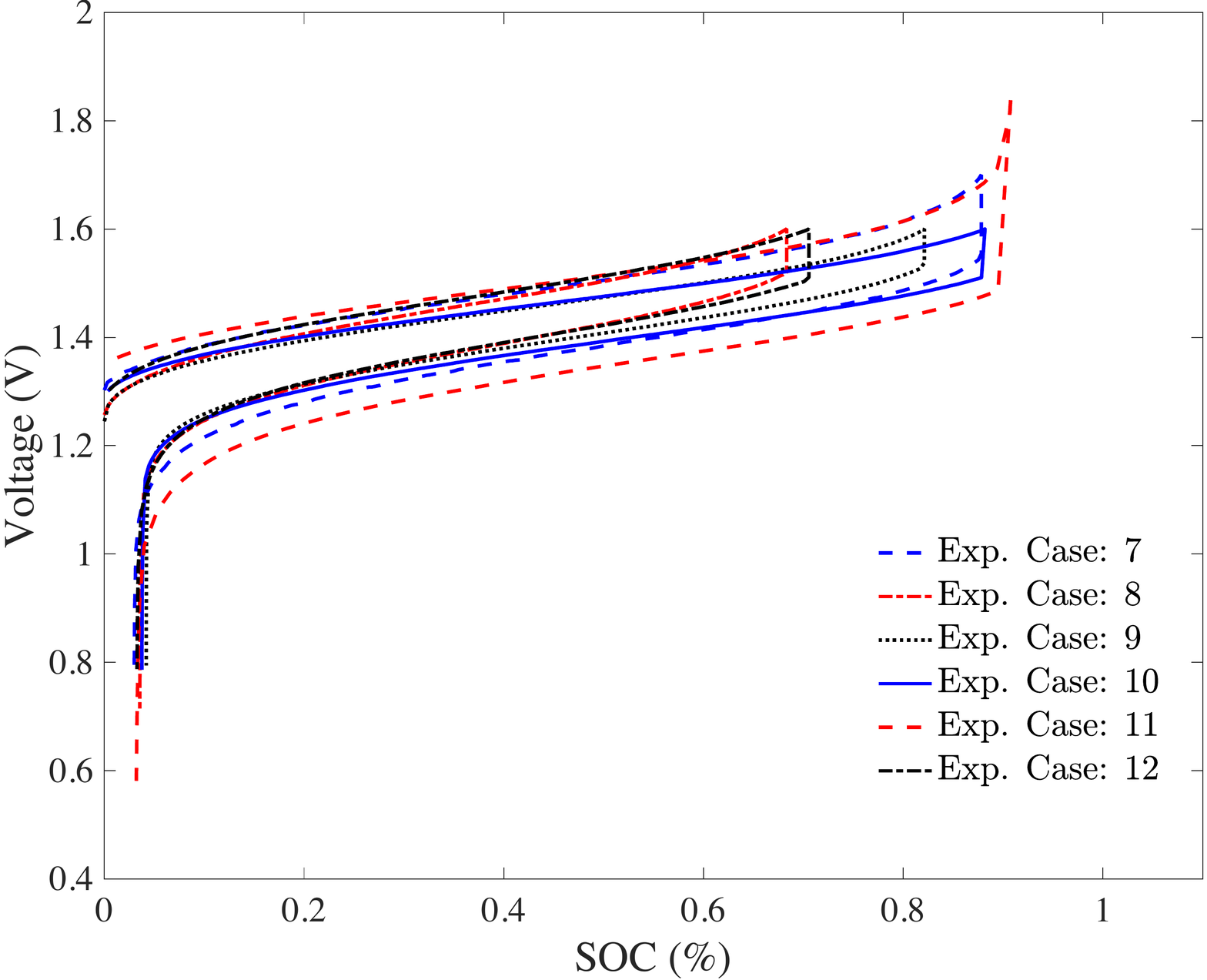}} \\
	%	\subfloat[] {\includegraphics[angle=90,width=2.5in]{results/ex1_validation/plot_Voltage_Eapen_mul_case_3}}
%	\subfloat[] {\includegraphics[width=2.5in]{results/PNNL_exp/exp_data}}
	\caption{The measured charge-discharge curves (third cycle) for the 12 PNNL experiments given in Table \ref{table:PNNL_exp}.}
	\label{fig:exp_raw_data}
\end{figure}

The cell voltage $E$ as a function of time (and converted to SOC using the relation in  \eqref{eq:soc_t}) was measured during each charge-discharge cycle.
The experimental results show that the first two charge-discharge cycles may exhibit poor Coulombic efficiency. Thus, we train the PCDNN model using measurements collected during the third cycle. The SOC-V data for the third cycle are plotted in Fig. \ref{fig:exp_raw_data}. 
In Section \ref{sec:res_exp}, we use the experimental data  to demonstrate the accuracy of PCDNN approach.
% In Section \ref{sec:res_exp}, we discuss the training and validation of the the PCDNN model using this experimental data.

%%% ---------------------------------
%%% ---------------------------------
\subsection{Data availability}
The data that support the findings of this study are available from the corresponding author upon request.

\section{Results}\label{sec:result}

As mentioned above, the values of the parameters $\sigma_e$, $S$, $k_n$, and $k_p$  vary significantly in different studies. %\cite{bhattacharjee2018precision,Lee2018,Lee2019c,Choi2020,cheng2020data}.
% Numeric reason
In addition to dependence on the operating conditions, a reason for this variation is the ill-posed nature of multiple-parameter estimation problems that often do not have a unique solution without proper regularization.
% may depend on the applied regularization.
% physical reason
%Another reason is that some lumped parameters, such as reaction rates, intrinsically depend on the operation conditions and the scale of the experiment, and therefore, they change in different experiments \cite{Wang2018i}.
%However, performing parameter calibration repeatedly for each experiment will be tedious and lack of generalization.
In Section \ref{sec:validation_0D}, we use synthetic data simulated with the 0D model to demonstrate that the PCDNN method provides necessary regularization to obtain accurate estimates of parameters when the parameter dependence on the operating conditions is not pronounced. 

In Section \ref{sec:res_exp}, we use the PCDNN method to learn parameters as functions of the operating conditions using the experimental data described in Section \ref{sec:data_pnnl}.  

%All numerical simulations are performed in a Mac OS environment with Intel® CoreTM i5-8500 (Hexa-Core, 3.00 GHz) processor and random-access memory of 32 GB. The detailed comparison of consumed computational resources for each polarization point between the two different models are listed in Table 2.

%%% ---------------------------------
% Example I: Validation with Eapen's data
%%% ---------------------------------
\subsection{Parameter estimation for simulation data}\label{sec:validation_0D}

%% --------  model data --------------
%%% ---------------------------------
\begin{table}[htb]
	\centering
	\small{
		\caption{Model parameters for the simulation data of the VRFB cell \cite{Eapen2019}.}
		\label{table:ex1_model}
		%	\resizebox{\columnwidth}{!}{}
		%	 \resizebox{1.0\textwidth}{!}{\begin{minipage}{\textwidth}
		%\end{minipage}}
		\scalebox{0.9}{
			\begin{tabular}{cccc}
				\toprule
				Symbol & Description  & Unit   &  Values   \\
				\hline
				$E_{p}^0$     & Standard equilibrium potential (Positive)   & $\text{V}$    & $1.004$   \\		
				$E_{n}^0$     & Standard equilibrium potential (Negative)   & $\text{V}$    & $-0.26$   \\	
				$n_{d}$     & Drag coefficient  & -    & $2.5$   \\	
				$k_p$      & Standard rate constant at $303$ K (Positive)   & $\text{m}$ $\text{s}^{-1}$   & $1.114 \times 10^{-4}$    \\	
				$k_n$      & Standard rate constant at $303$ K (Negative) & $\text{m}$ $\text{s}^{-1}$   & $1.798 \times 10^{-5}$   \\
				$S$     & Specific surface area & $\text{m}^{-1}$  & $420$ \\	
				$\epsilon$     & Porosity & - & $0.67$ \\
				$\sigma_{e}$     & Electrode conductivity & $\text{S}$ $\text{m}^{-1}$ & $1000$ \\	
				$\sigma_{c}$     & Current collector conductivity & $\text{S}$ $\text{m}^{-1}$ & $9.1 \times 10^{4}$ \\
				$T_{ref}$     & Reference temperature & $K$ & $293$ \\	
				\hline
				%			$h_e$     & Electrode length  & $\text{m}$   & $0.05$   \\	
				%			$b_e$     & Electrode thickness  & $\text{m}$   & $0.05$   \\
				$A_e$     & Electrode area  & $\text{m}^{2}$   & $0.0025$   \\
				$w_e$     & Electrode width  & $\text{m}$   & $0.003$   \\
				$w_m$    & Membrane width  & $\text{m}$   & $1.25 \times 10^{-4}$    \\
				$w_c$   & Current collector width  & $\text{m}$   & $0.015$    \\
				$V_r$     & Reservoir volume  & $\text{m}^{3}$   & $1 \times 10^{-4}$     \\			
				\bottomrule
			\end{tabular}%
		}
	}
\end{table}

% k+ is usually given from experimental resuls (Cheng 2020 [99]). k- is numerically fitted (smaller value)
\begin{table}[htb]
	\centering
	\small{
		\caption{Operating conditions for the simulation data of the VRFB cell \cite{Eapen2019}.}
		\label{table:ex1_condition}
		%	\resizebox{\columnwidth}{!}{}
		%	 \resizebox{1.0\textwidth}{!}{\begin{minipage}{\textwidth}
		%\end{minipage}}
		\scalebox{0.9}{
			\begin{tabular}{ccccc}
				\toprule
				Symbol 		&  Description  &  Unit		&  Reference values & Test range [Min, Max]  \\
				\hline
				$ \omega$ & Volumetric flow rate
				& $\text{m}^{3}$ $ \text{s}^{-1}$ & $4.17\times 10^{-7}$ & -  \\	
				$ j $ & Current density
				& $ \text{A}$ $\text{m}^{-2} $ & $300$ & [200,600]  \\	
				$c_{V}^0$ 					 & Initial vanadium concentration 
				& mol $\text{m}^{-3}$ & $500$ & -   \\		
				$ c_{\text{H}^+_n}^0$ & Initial $\text{H}^+$ concentration (negative)
				& mol $\text{m}^{-3}$ & $6000$  & -  \\
				$ c_{\text{H}^+_p}^0$ & Initial $\text{H}^+$ concentration (positive)
				& mol $\text{m}^{-3}$ & $6000$  & - \\					
				$ c_{\text{H}_2 \text{O}}$ & Initial $\text{H}_2 \text{O}$ concentration
				& mol $\text{m}^{-3}$ & $4.6 \times 10^{4}$ & -  \\			
				$ T$ & Temperature & $K$ & $303$ & - \\			
				\bottomrule
			\end{tabular}%
		}
	}
\end{table}

\begin{figure}[htb!]
	\centering
	%	\subfloat[] {\includegraphics[angle=90,width=2.0in]{results/ex1_validation/plot_Voltage_Eapen_mul_case_1}}
	\includegraphics[angle=0,width=3in]{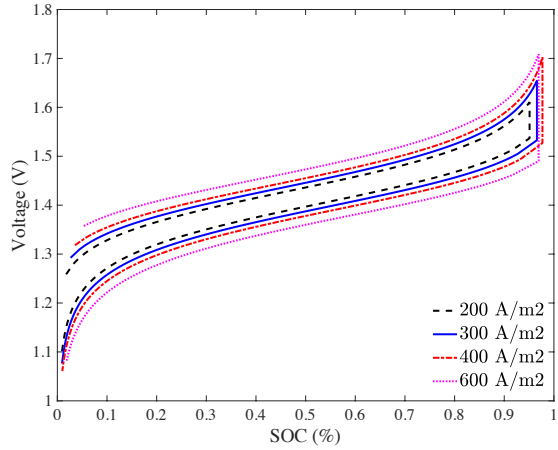}
	\caption{The SOC-V curves for different applied current densities ($j = 200$, $300$, $400$, $600$ $\text{A}/\text{m}^2$)  simulated by the ground truth 0D VRFB model. The numbers of data points for these four cases are 494,  494, 492, and 470, respectively.}
	\label{fig:data_0D}
\end{figure}
% ---Comment ----
% We first examine the accuracy of the proposed PINN method for parameter estimation using a synthetic data set generated with 0D model and parameters given in  \cite{Eapen2019} and summarized in Table \ref{table:ex1_model}.
% We use the 0D model with these parameters as the \textit{baseline} model.
% We generate four SOC-V curves for different current densities and operating conditions listed in Table \ref{table:ex1_condition}, as shown in Fig. \ref{fig:data_0D}.
% %These simulation data will be used as datasets to perform parameter identification, such that the estimated parameters can be validated by the given ground truth parameter. 
% The verification study using this dataset is given in Section \ref{sec:validation_0D}.
% ------ ----

To examine the accuracy of the proposed method for parameter estimation, we first test PCDNN using a simulation dataset consisting of four charge-discharge curves (see Fig. \ref{fig:data_0D}) that are generated by the 0D VRFB model
with the current densities $j=200$, 300, 400, and $600  \text{A}/\text{m}^{2}$. The 0D model parameters are 
 taken from \cite{Eapen2019} and listed in Table \ref{table:ex1_model}, and the operating conditions are given in Table \ref{table:ex1_condition}. %Thus, in this example, in the vector $\vec{x}=\{\tilde{u}, I, c^0_V\}$, only $I$ is changing in different. 
The parameter values $\vec{\mu} = (S, k_n, k_p, \sigma_e)$ given in Table \ref{table:ex1_model} are treated as ground truth and are used to simulate synthetic data (i.e., the ground truth values of $E^{cell}$).  
The simulation data are used to perform parameter identification, and the  estimated parameters are then compared against the ground truth parameters to validate the PCDNN method.

%%% ---------------------------------
%  Test for Simulation data
%%% ---------------------------------
% The simulation data obtained by the baseline 0D model (see Fig. \ref{fig:data_0D}) is used to verify the effectiveness of the proposed PINN approach for parameter estimation.
We use the SOC-V curves corresponding to $j=200$ and $400 \text{A}/\text{m}^{2}$ as a training set and the rest of the dataset to test the model. To demonstrate that the PCDNN is not very sensitive to the choice of the baseline parameters $\vec{\mu}_0$ with respect to the ground truth,  we select the baseline parameter values $S=1\times 10^3$ m$^{-1}$, $k_n= 5 \times 10^{-5}$ $\text{m}\text{s}^{-1}$, $k_p = 1 \times 10^{-4}$ $\text{m}\text{s}^{-1}$, and $\sigma_e = 0.5 \times 10^3$ $\text{S}\text{m}^{-1}$ that are different but within the same order of magnitude of the ground truth parameter values. 
% The units for these parameters are referred to Table \ref{table:para_model_PNNL}.

% We first consider $S$, $k_n$,  $k_p$, and $\sigma_e$ as the identifiable model parameters $\vec{\mu}$,
% and the uncalibrated reference parameters $\vec{\mu}^0$ are: $S=1 \times 10^3$, $k_n = 5 \times 10^{-5}$, $k_p = 1 \times 10^{-4}$, and $\sigma_e = 0.5 \times 10^3 $.
% ($\vec{\mu} = (S, k_n, k_p, \sigma_e)$)
%With considering the SOV-V curves obtained by the baseline 0D model (see Fig. \ref{fig:data_0D}) as the given database, the objective of this study is to  we applied the PINN approach to parameter estimation of a few  parameters used in the numerical model for the VRFB cell, where the cases of 
%Because we assume the parameters are constant, we also compare with the standard optimization.

\begin{table}[htb]
	\centering
	\small
	\renewcommand{\arraystretch}{0.7}
	\caption{The RMSE and estimated parameters obtained by PCDNN using different DNN sizes. The results given by the baseline and the least square estimation (LSE) are also provided as comparison. RMSE is evaluated on the test dataset. The ground truth parameters are: $S=4.2 \times 10^2$, $k_n = 1.798 \times 10^{-5}$, $k_p = 1.114 \times 10^{-4}$, and $\sigma_e = 1.000 \times 10^3$.}
% 	$S=4.2 \times 10^2$ $\text{m}^{-1}$, $k_n = 1.798 \times 10^{-5}$ $\text{m}\text{s}^{-1}$, $k_p = 1.114 \times 10^{-4}$ $\text{m}\text{s}^{-1}$, and $\sigma_e = 1.000 \times 10^3$ $\text{S}\text{m}^{-1}$.
	%	\resizebox{\columnwidth}{!}{%}
	\begin{tabular}{cccccc}
		\toprule
		DNN size & RMSE & $S$ [$\text{m}^{-1}$]  & $k_n$ [$\text{m}$ $\text{s}^{-1}$] & $k_p$ [$\text{m}$ $\text{s}^{-1}$] & $\sigma_e$ [$\text{S}$ $\text{m}^{-1}$]  \\
		\hline
		\multicolumn{1}{c}{$2 \times 20$} &  $1.555 \times 10^{-7}$ & $4.1301 \times 10^{2}$   &  $1.8285 \times 10^{-5}$   &  $1.1334 \times 10^{-4}$  & $1.0000 \times 10^{3}$  \\
		\multicolumn{1}{c}{$2 \times 30$} &  $0.626 \times 10^{-7}$ & $4.1323 \times 10^{2}$   &  $1.8275 \times 10^{-5}$   &  $1.1328 \times 10^{-4}$  & $1.0000 \times 10^{3}$ \\	
		\multicolumn{1}{c}{$2 \times 40$} & $2.256 \times 10^{-7}$ & $4.1370 \times 10^{2}$   &  $1.8254 \times 10^{-5}$   &  $1.1315 \times 10^{-4}$  & $1.0001 \times 10^{3}$ \\	
		\multicolumn{1}{c}{$3 \times 20$} & $0.784 \times 10^{-7}$ & $4.1570 \times 10^{2}$   &  $1.8167 \times 10^{-5}$   &  $1.1260 \times 10^{-4}$  & $1.0000 \times 10^{3}$ \\	
		\multicolumn{1}{c}{$3 \times 30$} &$1.720 \times 10^{-7}$  & $4.1663 \times 10^{2}$    &  $1.8126 \times 10^{-5}$   &  $1.1236 \times 10^{-4}$  & $1.0001 \times 10^{3}$\\	
		\multicolumn{1}{c}{$3 \times 40$} &$2.203 \times 10^{-7}$  & $4.1188 \times 10^{2}$    &  $1.8335 \times 10^{-5}$   &  $1.1364 \times 10^{-4}$  & $0.9998 \times 10^{3}$\\
		\hline
		Baseline                 & $2.477 \times 10^{-2}$  & $1.000 \times 10^{3}$    &  $5.000 \times 10^{-5}$   &  $1.000 \times 10^{-4}$  & $0.500 \times 10^{3}$ \\
		LSE                 & $1.244 \times 10^{-7}$  & $5.2201 \times 10^{2}$    &  $1.4467 \times 10^{-5}$   &  $0.89671 \times 10^{-4}$  & $1.0001 \times 10^{3}$ \\
		\bottomrule
	\end{tabular}
	\label{table:res_0D_PINN_p4_NN}
\end{table}

The structure of DNNs for $\hat{\mu}_i$ in Eq. \eqref{eq:mu_dnn_scale} is denoted as $n_l \times m_l$, where $n_l$ is the number of hidden layers and $m_l$ is the number of neurons per layer. More information about setting the DNN can be found in \ref{sec:appendix-DNN-appr}.
Because the size of the training data is relatively small ($N = 986$), the L-BFGS-B optimizer is adopted to train the PCDNN model. %The trained PCDNN model is used to predict the cell voltage for the test cases with $j=300$ and $600$ $ \text{A}/\text{m}^{2}$. 
The accuracy of the PCDNN prediction is given in terms of the root-mean-square error (RMSE) of the voltage prediction with respect to the reference test data.

%% Merge the following two paragraphs
% Table \ref{table:res_0D_PINN_p4_NN} summarizes the RMSE and the four estimated parameters by using a PCDNN with different DNN sizes.
% It shows that for all considered DNN sizes, the PCDNN yields accurate cell voltage predictions, as evidenced by the RMSE on the order of $10^{-7}$.
% On the other hand, the 0D model prediction with the baseline parameters $\vec{\mu}^0$ yields RMSE = $2.477 \times 10^{-2}$.

% It can be seen from Table \ref{table:res_0D_PINN_p4_NN} that the PCDNN predictions are practically independent on the DNN sizes, with the $3 \times 30$ DNNs giving the best estimated parameters. In the rest of the paper, we fix the DNN size to be $3 \times 30$ unless stated otherwise.
% In addition, the numerical results show that the parameters estimated by PCDNNs remain unchanged for different current density.
% This independence is expected because the synthetic data came from experiments that have the same parameter values. 

Table \ref{table:res_0D_PINN_p4_NN} summarizes the RMSE and the four estimated parameters as functions of the  DNN size.
It shows that for all considered DNN sizes, the PCDNN method yields accurate cell voltage predictions, as evidenced by the RMSE on the order of $10^{-7}$.
On the other hand, the 0D model prediction with the baseline parameters $\vec{\mu}^0$ yields RMSE = $2.477 \times 10^{-2}$.
The PCDNN predictions are practically independent of the DNN sizes, with the $3 \times 30$ DNNs giving the best estimated parameters. In the rest of the paper, we fix the DNN size to be $3 \times 30$ unless stated otherwise.

In addition, the numerical results show that the parameters estimated by PCDNN remain unchanged for different current density.
This independence is expected because the synthetic data came from experiments that use  operation-condition-independent parameter values. 

\begin{figure}[htb!]
	\centering
	\includegraphics[angle=0,width=5in]{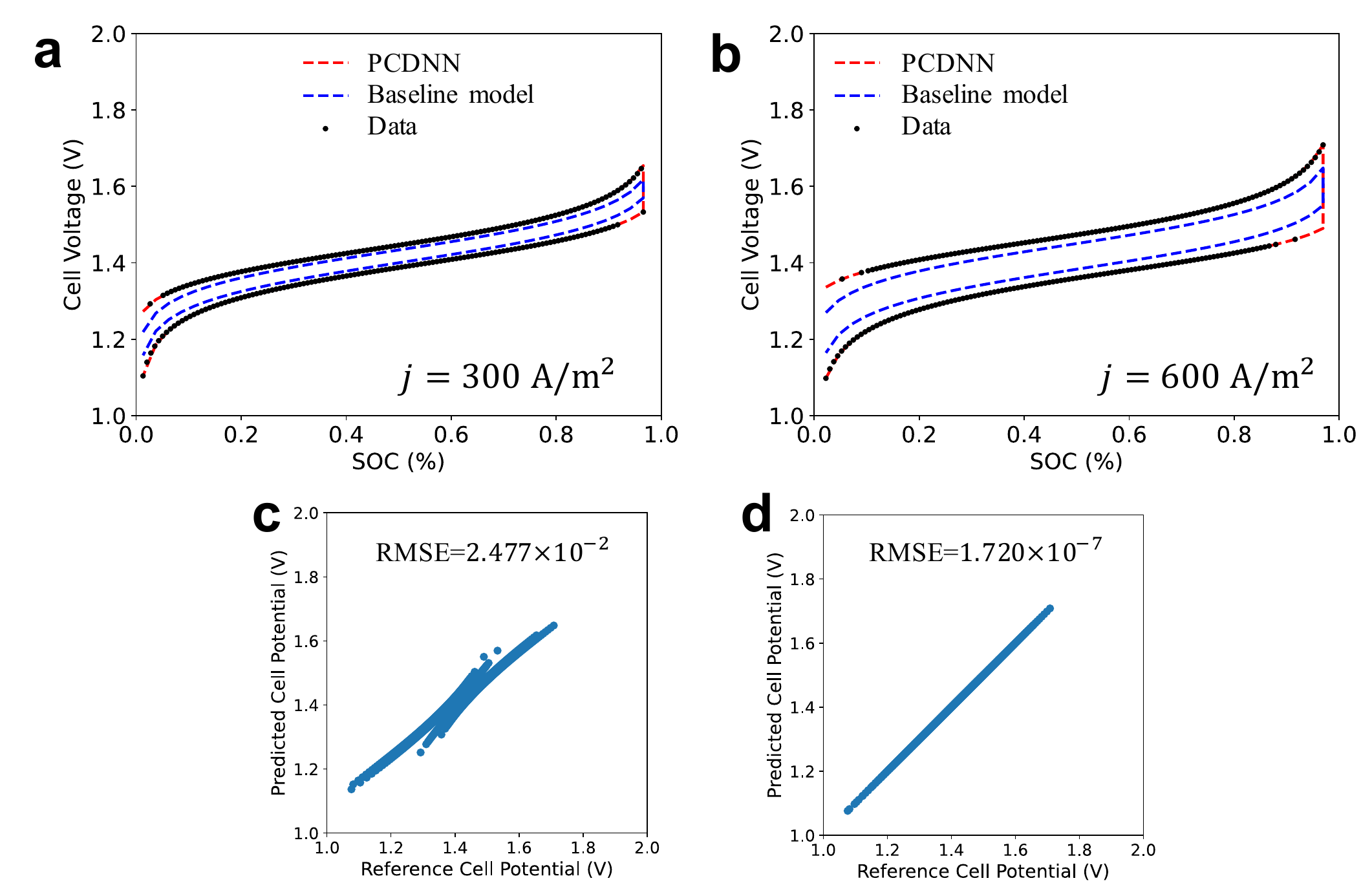}
	\caption{The comparison of cell voltage predictions obtained by the trained PCDNN model and the baseline 0D model with $\vec{\mu}^0$. The predicted SOC-V curves for the test cases $j=300$ $\text{A}/\text{m}^2$ and  $j=600$ $\text{A}/\text{m}^2$ are depicted in (a) and (b), respectively.
% 	where the reference data from the baseline model is also proived (black dots).
	(c) "one-one" plot to compare the voltage prediction by the baseline 0D model against the test data set. (d) "one-one" plot to compare the voltage prediction by the trained PCDNN model against the test data set. The DNN structure $3 \times 30$ is used for the PCDNN model, and the estimated parameters are shown in Table \ref{table:res_0D_PINN_p4_NN}.}
	%	The initial model parameters $\vec{\mu}^0$ are: $S=1 \times 10^3$, $k_n = 5 \times 10^{-5}$, $k_p = 1 \times 10^{-4}$, and $\sigma_e = 0.5 \times 10^3 $. The calibrated paramters by PINN $\hat{\vec{\mu}}$ are: $S=4.211 \times 10^2$, $k_n = 1.793 \times 10^{-5}$, $k_p = 1.112 \times 10^{-4}$, and $\sigma_e = 1.000 \times 10^3$.
	\label{fig:plot_0D_PINN_p4}
\end{figure}

Because we know that the ground truth parameters are the same for all cases in the simulation data, we can consider the standard least square estimation (LSE) approach for comparison \cite{Rahimian2011,Jokar2016}.
% Learning and Adaptation for Optimization and Control of Complex Renewable Energy Systems
% cursive Least Squares (RLS) and Kalman filter are generally applied to identify the parameters of ECM [32–36]. Since the traditional Recursive Least Squares (RLS) estimation is extremely sensitive to the noise, the parameters in the ECM may fail to converge to their true values under the measurement noise.
In the LSE approach, the parameters are found by solving the least square minimization problem,
\begin{equation}
	\min_{\{\mu_i\}_{i=1}^m} \frac{1}{N} \sum_{n=1}^{N}  [E^{cell}_{\mathcal{M}}(\{\mu_i\}_{i=1}^m;\vec{z}_n) - E^*(\vec{z}_n)]^2.
\end{equation}
Table \ref{table:res_0D_PINN_p4_NN} shows that the LSE approach identifies a set of parameters that results in an accurate voltage prediction with RMSE = $2.203 \times 10^{-7}$, which is of the same order as RMSE in the PCDNN method.
However, the parameters estimated by PCDNN are much closer to the ground truth parameters
(see Table \ref{table:res_0D_PINN_p4_NN}) than those from the LSE method, suggesting an improvement in PCDNNs over the standard LSE approach for this multiple parameter estimation problem. 
% that is nearly ill-posed.
%\textcolor{red}{This could be attributed to the DNN representation for parameters in Eq. \eqref{eq:mu_dnn_scale} that helps to mitigate the scaling issue and enhance the robustness for ill-conditioned problems.}
%\todo{This explanation is confusing. I thought that you learn log of parameters to deal with "the scaling issue." Do you also estimate $y=log\mu$ in the LSE approach? Also, in this example, the parameters are constants so you do not really need a DNN to represent a constant, right? QH: In LSE, I only use the same normalization. 2. Yes, the parameters are constants. I will double checked for this pts}
% Further analysis of the LSE approach is presented in the next subsections.
% Further comparison between the PCDNN and LSE approaches is presented in the next subsections.

Cell voltages predicted with the PCDNN model (with the DNN size $3 \times 30$)
for $j=300$ and $600$ $\text{A}/\text{m}^2$ are shown in Fig. \ref{fig:plot_0D_PINN_p4}a and b, respectively.
Fig. \ref{fig:plot_0D_PINN_p4}c and d present the "one-one" plots of the baseline 0D and PCDNN models predictions of $E^{cell}$ against the test dataset values of $E^{cell}$, respectively. 
Fig. \ref{fig:plot_0D_PINN_p4} shows that the PCDNN approach substantially improves the voltage prediction for different current density cases relative to the baseline 0D model. 

%%% ---------------------------------
%%% ---------------------------------
%\subsubsection{Dependence of LSE and PCDNN on the number of unknown parameters}
Next, we study the sensitivity of the PCDNN and LSE methods with respect to the number of unknown parameters. 
%Table \ref{table:res_0D_PINN_p4_NN} also demonstrates that the estimation of $S$, $k_n$, and $k_p$ parameters for the considered data is not unique, i.e., significantly different parameter values predicted by the PCDNN and LSE methods yield almost exact voltage predictions.
%In the reminder of this section, we investigate the accuracy of the LSE and PCDNN methods as a function of the number of parameters. or less number of unknown parameters.
 We do this by considering a case where only $S$, $k_n$, and $\sigma_e$ as unknown, while $k_p$ is set to its ground truth value $1.114 \times 10^{-4}$.
The PCDNN approach estimates the three parameters as  $S=4.1725 \times 10^2$, $k_n = 1.8097 \times 10^{-5}$, and $\sigma_e = 1.0002 \times 10^3$, which are close to the values that these parameters estimated in Table \ref{table:res_0D_PINN_p4_NN} where the number of unknown parameters was set to 4.
On the other hand, the LSE approach yields the vales $S=4.2179 \times 10^2$, $k_n = 1.7904 \times 10^{-5}$, and $\sigma_e = 0.9998 \times 10^3$, which are significantly different from the values of these parameters in Table \ref{table:res_0D_PINN_p4_NN} and closer to the ground truth values.

The inverse problem with three unknown parameters is significantly simpler than the one with four unknown parameters because of a nonlinear dependence between $k_p$ and $k_n$ in \eqref{eq:eta_act} for the activation overpotential.
%strong competitive effects, Competitively interative, strong interactive effect
These results show that the proposed PCDNN approach outperforms the standard LSE method by being less sensitive to the number of unknown parameters and a nonlinear dependence between the unknown parameters.

It must be pointed out that in this case, the values of model parameters that need to be learned by the DNNs are constant and fixed to their ground truth values.
%is worth pointing out that although each parameter is modeled as a DNN function of input (operation) conditions, the PCDNN method is able to correctly learn the constant values of parameters independent on the operation conditions in this validation example. % and  the trained PCDNN model is able to recover the constant functions for the estimated parameters, which is physically consistent to the fact that the baseline model indeed generates the simulation data with the constant parameters in Table \ref{table:ex1_model}.
%
We also note that with both PCDNN and LSE, 
the parameter $\sigma_{e}$ is estimated more accurately than the other parameters. This is because the dependence of $E^{cell}$ on $\sigma_{e}$ is simpler than on the rest of the parameters.
%, i.e., independently used to determine the ohmic loss without interacting with other unknown parameters, thus, it is relatively easier to identify.

%%% ---------------------------------
% Example: II
%%% ---------------------------------
\subsection{Learning parameters as functions of operating conditions from experimental data}\label{sec:res_exp}
Here, we consider the 12 experiments described in Section \ref{sec:data_pnnl}, with the operating conditions and the known (measured) VRFB parameters summarized in Table \ref{table:PNNL_exp}.
%This table shows that the species concentration ($c_i^0$), applied current ($I$), flow rate ($\omega$), tank volume ($V_r$), and the type of membrane (i.e., thickness $w_m$) are different across these 12 experiments.
In all considered experiments, $A_e = 20$ $\text{cm}^{2}$, $w_e = 0.4$ $\text{cm}$, $w_c = 1.5$ $\text{cm}$, and temperature are assumed to be constant $T = 298$ $K$. 
All other parameters that are required for the 0D model of these experiments are given in Table \ref{table:ex1_model}.

% The reference values \textcolor{red}{$\vec{\mu}^0$} of the model parameters to be estimated are given in Table \ref{table:para_model_PNNL}.
The baseline  values $\vec{\mu}^0$ of the unknown 0D model parameters are taken from \cite{Chen2021} and listed in Table \ref{table:para_model_PNNL}. This table also lists 
a physically admissible range for each unknown parameter $\mu_i$, where the upper and lower bounds are estimated according to \cite{Choi2020} based on the previous studies reported in  \cite{Shah2008,Shah2011,Sharma2014,Eapen2019,Lee2019c,Chen2021}.

Given the large number of operating conditions (Table \ref{table:PNNL_exp}), %and uncertainties  due to processes that are not considered in the 0D model, e.g.,  the temperature fluctuations, 
$E^{cell}$ varies significantly in the experiments, as shown in Fig. \ref{fig:exp_raw_data}. 
It is neither practical nor very useful to perform parameter fitting for each experiment because of the ill-posed nature of the parameter estimation problem. Thus, the proposed PCDNN method is used to learn the model parameters based on the ensemble of experimental data corresponding to the various operating conditions.

\begin{table}[htb]
	\centering
	\small{
		\caption{The baseline values and feasible ranges of the model parameters to be estimated from experiments.}
		\label{table:para_model_PNNL}
		\scalebox{0.8}{
			\begin{tabular}{ccccc}
				\toprule
				Symbol & Description  & Unit   &  Baseline values \cite{Chen2021}  & Test range [Min, Max] \cite{Choi2020}  \\
				\hline
				$S$          & Active surface area                                   & $\text{m}^{-1}$                    & $3.48 \times 10^{4}$   & $[1.62 \times 10^3, 1.62 \times 10^5]$ \\	
				$k_n$      & Standard rate constant at $T$ (Negative) & $\text{m}$ $\text{s}^{-1}$    & $5.0 \times 10^{-8}$   &  $[1.7 \times 10^{-8}, 6.8 \times 10^{-6}]$  \\
				$k_p$      & Standard rate constant at $T$ (Positive)   & $\text{m}$ $\text{s}^{-1}$   & $1.0 \times 10^{-7}$    &  $[1.7 \times 10^{-8}, 6.8 \times 10^{-6}]$\\	
				$\sigma_{e}$     & Electrode conductivity                     & $\text{S}$ $\text{m}^{-1}$    & $500$       &  $[1.0 \times 10^2, 1.0 \times 10^4]$ \\	
				%				\hline
				%				$A_e$     & Electrode area  & $\text{m}^{2}$   & $0.002$ & -   \\
				%				$w_e$     & Electrode width  & $\text{m}$   & $0.004$    & - \\
				%				$w_c$     & Current collector width  & $\text{m}$   & $0.015$  \\
				%				$T$         & Operating temperature   & $K$    & $298$   \\			
				\bottomrule
			\end{tabular}%
		}
	}
\end{table}

%%% ---------------------------------
% Exp 22: case study
%%% ---------------------------------
\subsubsection{Learning parameters from the experimental dataset}\label{sec:exp_case_study1}
Here, we randomly select 60\% data points from the 12 experimental SOC-V curves (see Fig. \ref{fig:exp_raw_data}) for training the PCDNN model.
%all the 12 experimental data sets are considered in an ensemble dataset, where we randomly select 60\% data points for training the PCDNN model.
The remaining 40\% data are used as the test data to evaluate the prediction performance. 
%We consider the parameter values calibrated for Experiment Case 6 as the reference values, as shown in Table \ref{table:para_model_PNNL}, and used as the initial guess $\mu_i^0$.
%The reference parameters shown in Table \ref{table:para_model_PNNL} are used as the initial guess $\mu_i^0$.
% We consider the parameter values calibrated in \cite{Chen2021} as the reference, as shown in Table \ref{table:para_model_PNNL}, and also used as the initial guess $\mu_i^0$.
After training the PCDNN model, we obtain the parameters $\hat{\mu}_i(\vec{x}; \theta_i)$ as functions of the operating conditions $\vec{x}=\{\tilde{u}, I, c^0_V\}$. Then, the PCDNN estimated model parameters for each experiment can be computed by evaluating $\hat{\mu}_i(\vec{x}; \theta_i)$ for the operating conditions $\vec{x}$ in the considered experiment. 
% The values of parameters computed for experiments 2, 4, 8, and 11 from the DNN models of these parameters are given in Table \ref{table:res_exp_PINN_case}.

% \todo{QH: delete Table 6.}
% % Full 4 case
% \begin{table}[htb]
% 	\centering
% 	\renewcommand{\arraystretch}{0.8}
% 	{
% 		\caption{The estimated parameters and the RMSE obtained by the PCDNN approach for the selected experiments.\textcolor{red}{QH: This table is to be deleted.}}
% 		\label{table:res_exp_PINN_case}
% 	}
% 	\resizebox{0.8\textwidth}{!}  % use \columnwidth
% 	{
% 		\begin{tabular}{ccccccc}
% 			\toprule
% 			\multicolumn{2}{c}{ Exp. ID} & RMSE & \multicolumn{4}{c}{PCDNN estimated parameters}  \\
% 			\hline
% % 			& &  & $S$ & $k_n$ & $k_p$ & $\sigma_e$  \\
% % 			\multicolumn{2}{c}{1} & $2.875 \times 10^{-2}$			& $3.62 \times 10^4$   &  $1.75 \times 10^{-7}$   &  $1.67 \times 10^{-7}$  & $1.56 \times 10^{2}$ \\
% 			\multicolumn{2}{c}{2} & $1.716 \times 10^{-2}$			& $3.62 \times 10^4$   &  $1.75 \times 10^{-7}$   &  $1.67 \times 10^{-7}$  & $1.56 \times 10^{2}$ \\			
% 			\multicolumn{2}{c}{4} & $6.331 \times 10^{-3}$			& $2.65 \times 10^4$   &  $1.70 \times 10^{-8}$   &  $2.08 \times 10^{-7}$  & $3.35 \times 10^{2}$ \\
% 			\multicolumn{2}{c}{8} & $3.602 \times 10^{-2}$ 			& $2.58 \times 10^4$   &  $1.91 \times 10^{-7}$   &  $1.80 \times 10^{-7}$  & $1.81 \times 10^{2}$ \\
% 			\multicolumn{2}{c}{11} & $6.403 \times 10^{-3}$				& $1.51 \times 10^4$   &  $1.51 \times 10^{-7}$   &  $2.31 \times 10^{-7}$  & $3.21 \times 10^{2}$ \\
% 			\bottomrule
% 		\end{tabular}
% 	}
% \end{table}

% Full 12 case
\begin{table}[htb]

	\centering
	\renewcommand{\arraystretch}{0.8}{\caption{The estimated parameters and the RMSE obtained by the PCDNN approach for the 12 experiments. The constant parameters estimated by LSE are also provided for comparison.}\label{table:res_exp_PINN_case}}
	
	\resizebox{1.0\textwidth}{!}{		\begin{tabular}{ccccccc}
			\toprule
% 			\multicolumn{2}{c}{ Exp. ID} & RMSE & \multicolumn{4}{c}{PCDNN estimated parameters}  \\
% 			\hline
% 			& &  & $S$ & $k_n$ & $k_p$ & $\sigma_e$  \\
			\multicolumn{2}{c}{ Exp. ID} & RMSE & $S$ & $k_n$ & $k_p$ & $\sigma_e$\\
			\hline
 			% \multirow{12}{*}{PCDNN} &
 		    & & & \multicolumn{4}{c}{PCDNN estimated parameters}
 		    \\
			\multicolumn{2}{c}{1} & $2.875 \times 10^{-2}$			& $2.95 \times 10^4$   &  $7.07 \times 10^{-8}$   &  $1.19 \times 10^{-7}$  & $1.00 \times 10^{2}$ \\
			\multicolumn{2}{c}{2} & $1.716 \times 10^{-2}$			& $3.62 \times 10^4$   &  $1.75 \times 10^{-7}$   &  $1.67 \times 10^{-7}$  & $1.56 \times 10^{2}$ \\	
			\multicolumn{2}{c}{3} & $7.620 \times 10^{-2}$			& $2.61 \times 10^4$   &  $1.14 \times 10^{-7}$   &  $1.90 \times 10^{-7}$  & $2.42 \times 10^{2}$ \\
			\multicolumn{2}{c}{4} & $6.331 \times 10^{-3}$			& $2.65 \times 10^4$   &  $1.70 \times 10^{-8}$   &  $2.08 \times 10^{-7}$  & $3.35 \times 10^{2}$ \\
			\multicolumn{2}{c}{5} & $9.822 \times 10^{-3}$			& $2.67 \times 10^4$   &  $1.70 \times 10^{-8}$   &  $2.13 \times 10^{-7}$  & $3.43 \times 10^{2}$ \\
			\multicolumn{2}{c}{6} & $7.141 \times 10^{-3}$			& $2.86 \times 10^4$   &  $1.04 \times 10^{-7}$   &  $2.56 \times 10^{-7}$  & $1.95 \times 10^{2}$ \\			
			\multicolumn{2}{c}{7} & $3.243 \times 10^{-2}$			& $2.61 \times 10^4$   &  $1.14 \times 10^{-7}$   &  $1.91 \times 10^{-7}$  & $2.42 \times 10^{2}$ \\
			\multicolumn{2}{c}{8} & $3.602 \times 10^{-2}$ 			& $2.58 \times 10^4$   &  $1.91 \times 10^{-7}$   &  $1.80 \times 10^{-7}$  & $1.81 \times 10^{2}$ \\
			\multicolumn{2}{c}{9} & $3.472 \times 10^{-2}$ 			& $2.58 \times 10^4$   &  $1.91 \times 10^{-7}$   &  $1.80 \times 10^{-7}$  & $1.81 \times 10^{2}$ \\			
			\multicolumn{2}{c}{10} & $3.137 \times 10^{-2}$ 			& $2.61 \times 10^4$   &  $1.14 \times 10^{-7}$   &  $1.91 \times 10^{-7}$  & $2.42 \times 10^{2}$ \\
			\multicolumn{2}{c}{11} & $6.403 \times 10^{-3}$				& $1.51 \times 10^4$   &  $1.51 \times 10^{-7}$   &  $2.31 \times 10^{-7}$  & $3.21 \times 10^{2}$ \\
			\multicolumn{2}{c}{12} & $3.836 \times 10^{-2}$				& $3.56 \times 10^4$   &  $1.38 \times 10^{-6}$   &  $1.46 \times 10^{-7}$  & $1.00 \times 10^{2}$ \\
% 			\multicolumn{2}{c}{PCDNN (all cases)} & $3.267 \times 10^{-2}$ & & & & \\
 			\hline
  		    \multicolumn{2}{c}{LSE (all cases)} &  & $3.04 \times 10^4$   &  $2.90 \times 10^{-8}$   &  $1.38 \times 10^{-7}$  & $5.68 \times 10^{2}$ \\
			\bottomrule
		\end{tabular}
		}
\end{table}

The estimated parameters of the PCDNN model for the whole dataset (i.e., all 12 experiments) are given in Table \ref{table:res_exp_PINN_case}.
% \todo{I would combine Tables 6 and 7. In the combined table, I would list the values of parameters for all 12 experiments and the LSE estimated parameters. This way you do not need to give PCDNN ranges. The baseline values I would list in the caption.QH: Done}
% As we can see, the ranges of the parameters in the PCDNN models for the considered experiments are substantially smaller than those reported in literature (see Table \ref{table:para_model_PNNL}).
% \todo{I assume that the  parameters ranges reported in Table 5 cover more a wider range of operating conditions than those in our 12 experiments. So, I am not sure what is the significance of this observation. QH: Deleted.}
%, which suggests that the PINN can identify the proper intrinsic parameters from the indirect measurements (cell voltage).
It can be seen that, among the four estimated parameters, $S$ has the smallest range of values. This can be explained by the fact that these experiments adopt the same carbon fiber electrodes, and the variations are mainly due to different flow rates that affect the spatial distribution of species in the electrodes.
% \todo{If you make this statement, then you should also explain what causes higher variations in the other parameters. QH: Page 4 line 35-44 in Introduction explained why generally reaction rates vary significantly in terms of flow rate and current. But for these specific 12 experiments, there could be other factors affecting the variations.}
%\textcolor{blue}{
%[To confirm with Litao]
%The dependence of the parameters $k_p$ and $k_n$ on operation conditions is in agreement with the study in \cite{Kroner2020}.%, we observed that both $k_p$ and $k_n$ (representing electrochemical reaction) depend on the operating condition, i.e., concentration and the state of charge, but it shows the standard rate constant $k_p$ is less sensitive than $k_n$.
% Not sure the statement.
%In fact, it has been reported that the $k_p$ can be determined experimentally \cite{Shah2008,Gattrell2004} while $k_n$ need to be numerically fitted, indicating $k_n$ with higher uncertainty over $k_p$.
%}
% $\sigma_e$ is also need to fitted. But
 
\begin{figure}[htb!]
	\centering
	\includegraphics[width=6in]{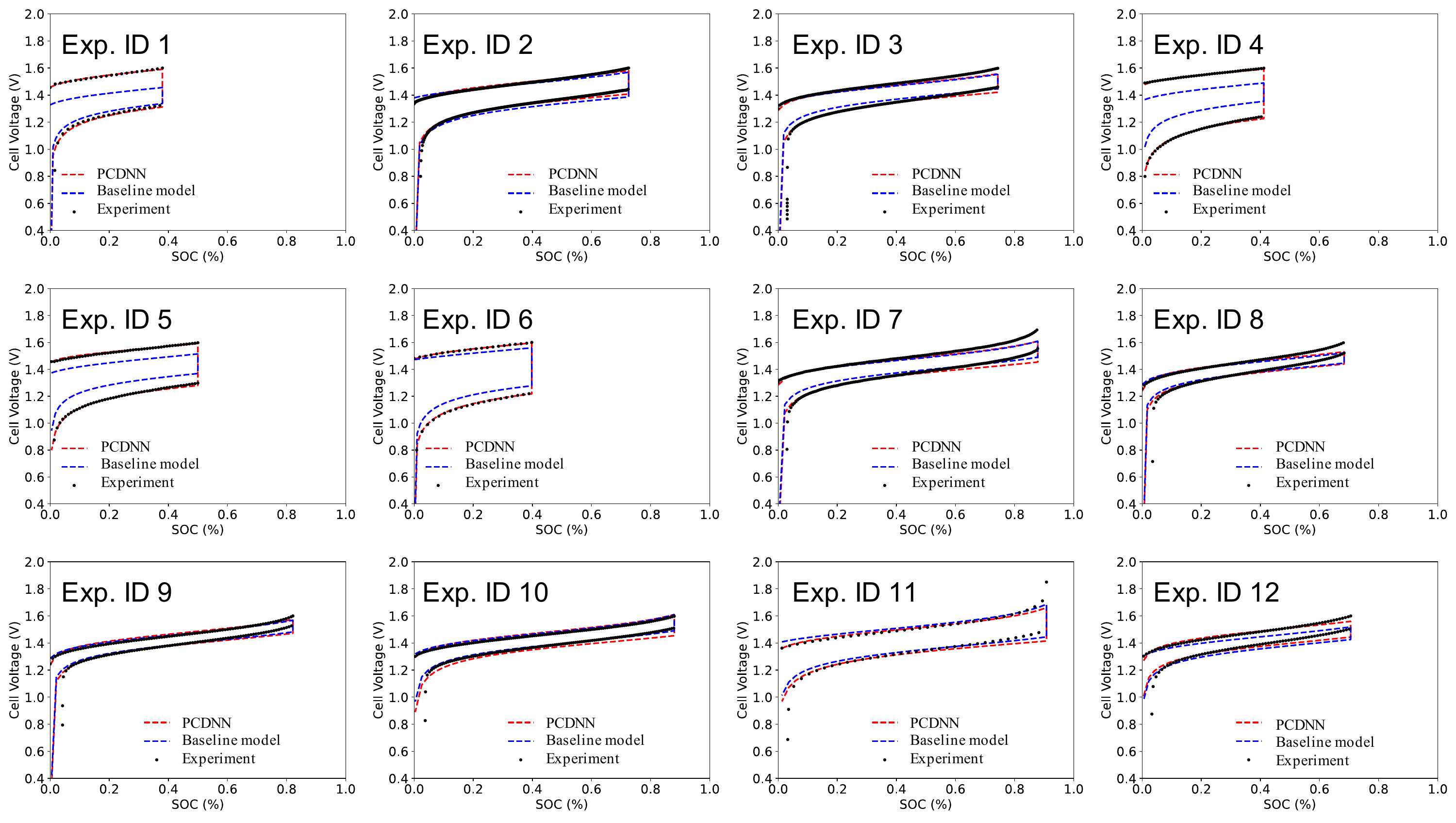}
	\caption{The comparison of SOC-V predictions for all 12 experiments by the PCDNN model (dashed red line) and the VRFB 0D model (dashed blue line) using the baseline parameters in Table \ref{table:para_model_PNNL}. The black dots denote the experimental data.}
    \label{fig:plot_exp_PINN_all}
\end{figure}

\begin{figure}[htb!]
	\centering
	\includegraphics[width=6in]{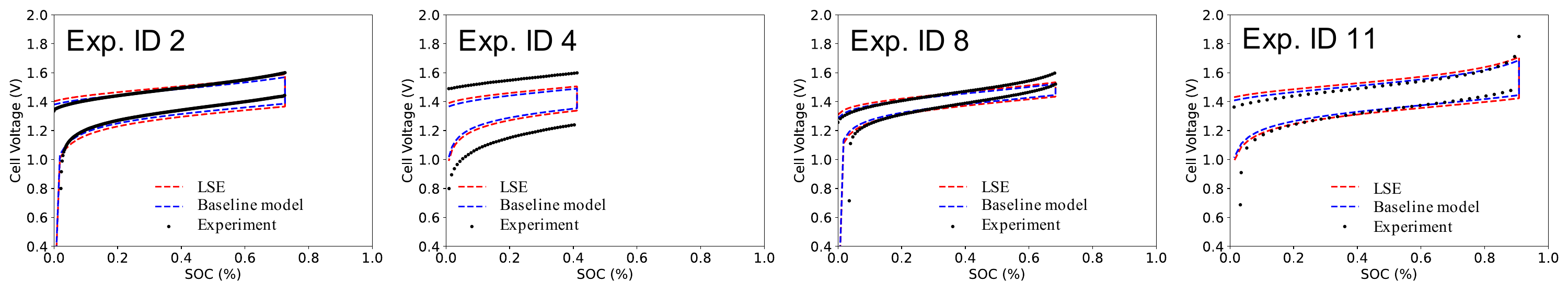}
	\caption{The comparison of SOC-V predictions for exemplary experiments by the VRFB 0D model using the LSE estimated parameters (dashed red line) in Table \ref{table:res_exp_PINN_case} and the baseline parameters (dashed blue line) in Table \ref{table:para_model_PNNL}. The black dots denote the experimental data.}
	\label{fig:plot_exp_LSE_case}
\end{figure}

%% RMSE 
% The RMSE of each experiment is shown in Table \ref{table:res_exp_PINN_case}, 
With the various estimated parameters, the RMSE of the PCDNN voltage prediction against the test data is $3.267 \times 10^{-2}$, which is about $35\%$ and $40\%$ smaller than those of the standard LSE approach and the baseline VRFB 0D model, respectively.
Fig. \ref{fig:plot_exp_PINN_all} shows that the PCDNN voltage prediction agrees well with the experimental data for experiments 1-6 (that use the Nafion 115 membrane) for the entire range of SOC values and is significantly better than the baseline 0D model predictions.
For experiments 7--12, there are some discrepancies between the PCDNN predictions and experimental data for very small and large SOC values.  These discrepancies are due to  simplifications in the 0D model and cannot be corrected through the choice of different values for the 0D model parameters. However, these results show that the proposed approach improves the prediction of the 0D model with constant parameters for all considered experiments. Also, it is important to note that using higher-dimensional flow models instead of the 0D model should further improve the predictive ability of the proposed approach. 
%This is because the adopted 0D model encoded in PCDNN is limited when it comes to describing the complex details in the real experiments. This suggests that a more sophisticated physical model should be encoded in the PCDNN model.

We also note that the standard LSE approach hardly improves the voltage prediction compared to the baseline 0D model using values from the literature as illustrated in Fig. \ref{fig:plot_exp_LSE_case} for experiments 2, 4, 8, and 11. %compares the LSE and uncalibrated 0D model predictions for  For example, the LSE voltage prediction for experiment 4 does not agree with the experimental data (see Fig. \ref{fig:plot_exp_LSE_case}).
%The SOC-V curves generated by the LSE approach for the experiment cases ID 2, 4, 8 and 11 are shown in Fig. \ref{fig:plot_exp_LSE_case}. 
%These results demonstrate the importance 

%%% ---------------------------------
% Exp 22: case study 2
%%% ---------------------------------
\subsubsection{Leave-one-out testing of the DNN parameter models}\label{sec:exp_case_study2}
%% Plot & RMSE for the comparison
\begin{figure}[htb!]
	\centering
	\includegraphics[width=0.8\textwidth]{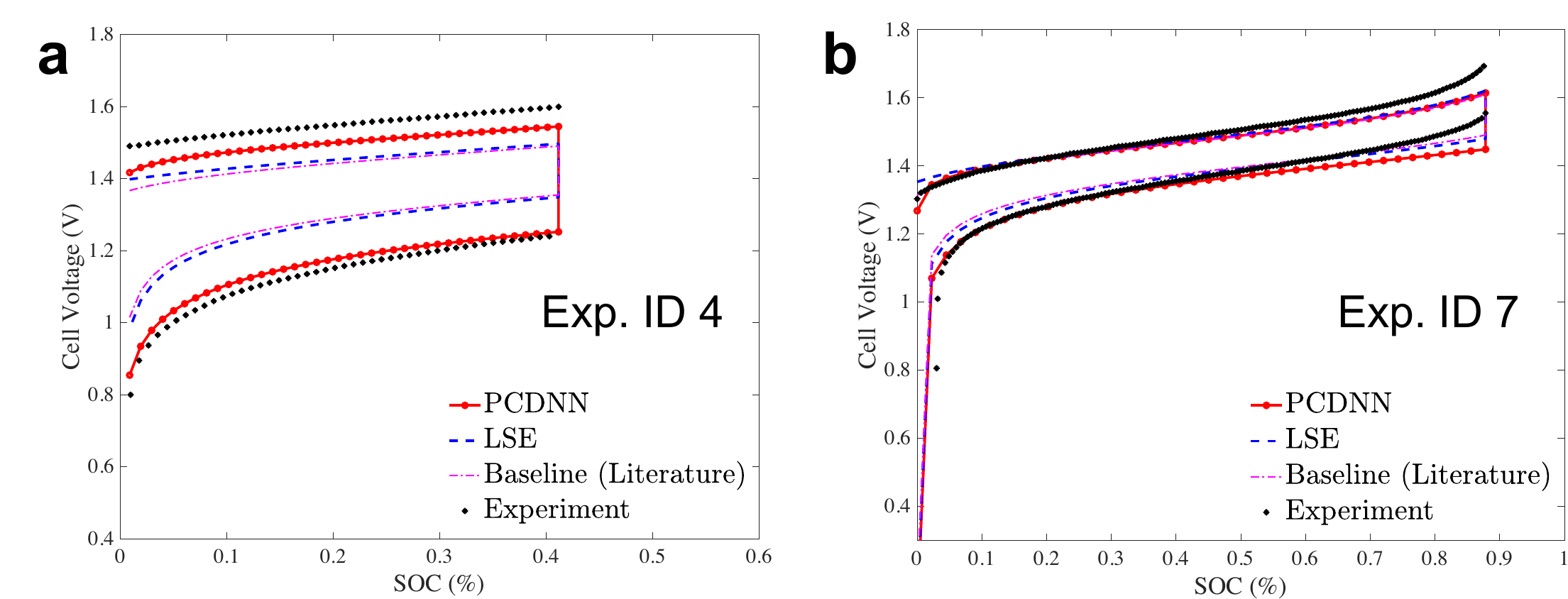}
	\caption{The comparison of SOC-V predictions for experiments (Cases 4 and 7) by using PCDNN and LSE parameter estimation approaches. 
% 	The results obtained by using the PCDNN and LSE estimated parameters are shown in (a) and (b), whereas that by LSE are shown in (c) and (d).
	The VRFB 0D model (dashed blue line) using the baseline parameters in Table \ref{table:para_model_PNNL} is also provided.}
	\label{fig:plot_exp_PINN_set2}
\end{figure}

\begin{table}[htb]
	\centering
	%	\small
	{
		\renewcommand{\arraystretch}{0.8}
		\caption{The comparison of the RMSE and estimated parameters by the PCDNN and LSE approaches for different experiment cases.}
		\label{table:res_exp_PINN_set2}
	}
	\resizebox{0.8 \textwidth}{!}  % use \columnwidth
	{
		\begin{tabular}{cccccc}
			\toprule
			& RMSE & $S$ & $k_n$ & $k_p$ & $\sigma_e$  \\
			\hline
			\multicolumn{2}{c}{Baseline model$^*$}  \\				
			Exp. ID 4 &  $1.271 \times 10^{-1}$ &  \multirow{2}{*}{ $3.48 \times 10^4$} &  \multirow{2}{*}{$5.00 \times 10^{-8}$}   &  \multirow{2}{*}{$1.00 \times 10^{-7}$}  & \multirow{2}{*}{$5.00 \times 10^{2}$} \\
			Exp. ID 7 &  $3.651 \times 10^{-2}$ \\
			\hline
			\multicolumn{2}{c}{ Least-square estimation}  \\			
			Exp. ID 4 &  $1.152 \times 10^{-1}$         &  \multirow{2}{*}{ $2.91 \times 10^4$} &  \multirow{2}{*}{$5.31 \times 10^{-8}$}   &  \multirow{2}{*}{$5.23 \times 10^{-8}$}  & \multirow{2}{*}{$1.29 \times 10^{3}$} \\
			Exp. ID 7 &  $3.274 \times 10^{-2}$ \\
			\hline
			\multicolumn{2}{c}{ PCDNN}  \\
			Exp. ID 4  &  $4.097 \times 10^{-2}$ & $3.01 \times 10^4$   &  $1.70 \times 10^{-8}$   &  $3.33 \times 10^{-7}$  & $1.00 \times 10^{4}$ \\
			Exp. ID 7  &  $3.357 \times 10^{-2}$ & $2.93 \times 10^4$   &  $2.42 \times 10^{-8}$   &  $5.30 \times 10^{-7}$  & $1.00 \times 10^{4}$ \\
			\bottomrule
		\end{tabular}
	}
	\\
	\small{\footnotesize{$^*$ use the baseline parameters in Table \ref{table:para_model_PNNL}}}
\end{table}

In this section, we investigate the effectiveness of the proposed PCDNN approach for modeling new or unseen experiments using the leave-one-out approach. 
That is, the PCDNN model is used to predict experiments with operating conditions different from those in the training dataset.
Figure \ref{fig:plot_exp_PINN_set2} and Table \ref{table:res_exp_PINN_set2} show a detailed analysis of the leave-one-out tests for experiments 4 and 7. Table \ref{table:res_exp_PINN_set2_all} gives RMSE in the leave-one-out tests for all 12 experiments. 

We select experiments 4 and 7 for the detailed analysis because they are performed under different operation conditions and use different membranes: the Nafion 115 membrane in experiment 4 and Nafion 212 in experiment 7.  
Figure \ref{fig:plot_exp_PINN_set2} shows the cell voltage for the test experiments as predicted by the PCDNN approach. 
Although no measurement data associated with the test experiments (Cases 4 and 7) are included in training the PCDNN model, the PCDNN prediction shows an improved agreement with the experimental data compared to the baseline 0D model.
The RMSEs of the PCDNN approach for experiments 4 and 7 are $4.097 \times 10^{-2}$ and $3.357 \times 10^{-2}$, respectively, which are smaller than the RMSEs $1.271 \times 10^{-1}$ and $3.651 \times 10^{-2}$ in the baseline 0D model.
Also, these results show that the LSE approach is less effective than PCDNN in predicting the voltage responses for the fourth experiment, as shown in Fig. \ref{fig:plot_exp_PINN_set2}a, where its RMSE is $1.152 \times 10^{-1}$.

\begin{table}[htb]
	\centering
	%	\small
	{
		\renewcommand{\arraystretch}{0.8}
		\caption{\textcolor{red}{RMSE in the voltage estimated with the 0D model with the PCDNN and LSE parameterazation for the 12 experiments in the leave-one-out test. For comparison, RMSE in the estimated voltage with the baseline 0D model is also shown.}}
		\label{table:res_exp_PINN_set2_all}
	}
	\resizebox{0.8 \textwidth}{!}  % use \columnwidth
	{
		\begin{tabular}{ccccc}
			\toprule
			Exp. ID & Baseline model$^*$ &  Least-square estimation & PCDNN   \\
			\hline
			1 & $1.036 \times 10^{-1}$  &  $9.035 \times 10^{-2}$ & $2.865 \times 10^{-2}$ \\ 
			2 & $2.641 \times 10^{-2}$ &  $3.508 \times 10^{-2}$ &  $1.682 \times 10^{-2}$      \\
			3 & $8.707 \times 10^{-2}$ &  $8.235 \times 10^{-2}$ & $7.552 \times 10^{-2}$\\
			4 & $1.271 \times 10^{-1}$ &  $1.152 \times 10^{-1}$ & $4.097 \times 10^{-2}$\\
			5 & $8.957 \times 10^{-2}$ &  $ 7.836 \times 10^{-2}$ & $9.115 \times 10^{-3}$\\
			6 & $5.367 \times 10^{-2}$ &  $ 4.500 \times 10^{-2}$ & $6.982 \times 10^{-3}$\\
			7 & $3.651 \times 10^{-2}$ &  $3.274 \times 10^{-2}$ & $3.357 \times 10^{-2}$\\
		    8 & $3.906 \times 10^{-2}$ &  $ 3.835 \times 10^{-2}$ & $3.751 \times 10^{-2}$\\
			9 & $3.590 \times 10^{-2}$ &  $ 3.609 \times 10^{-2}$ &$3.462 \times 10^{-2}$\\
			10 & $2.951 \times 10^{-2}$ &  $ 3.074 \times 10^{-2}$ &$3.310 \times 10^{-2}$\\
			11 & $6.988 \times 10^{-2}$ &  $ 6.758\times 10^{-2}$ &$6.336 \times 10^{-2}$\\
			12 & $4.979 \times 10^{-2}$ &  $ 4.880\times 10^{-2}$ &$3.943 \times 10^{-2}$\\
			\bottomrule
		\end{tabular}
	}
	\\
	\small{\footnotesize{$^*$ use the baseline parameters in Table \ref{table:para_model_PNNL}}}
\end{table}

\textcolor{red}{Table \ref{table:res_exp_PINN_set2_all} shows the comparison of the RMSE in voltage estimated with the 0D model with the PCDNN and LSE parameteraziations using the leave-one-out approach for all 12 experiments. For comparison, we also show the RMSE in the estimated voltage in the 0D model with the baseline parameters. We can see that even with a relatively small data set (only 11 experiments) the  PCDNN model is more accurate than the 0D model with parameters estimated using the LSE method for all but experiments 7 and 10. When compared to the 0D model with baseline parameters, the PCDNN is more accurate for all but the experiment 10. The applications of physics-constrained DNN methods to diffusion and advection-diffusion systems demonstrated that the accuracy of these methods improves with the increasing number of measurements \cite{he2020physics,Tartakovsky2020a}. Therefore, we expect that the accuracy of the PCDNN parameterization  versus the LSE paramerization  would increase as more data becomes available. }   

Like many other traditional parameter estimation methods, LSE finds the optimal set of parameters for a single instant of the operating conditions. %
%But it ignores the underlying differences of the measurements between different experiments, and cannot consider the possible model dependence on the operating conditions, which can be significant. 
The approach proposed here allows learning the functional relationships between the model parameters and operating conditions. We note that in general, functions are infinite dimensional, i.e., an infinite number of values are required to represent an unknown function, and using parameter estimation methods like LSE is poorly suited for learning functional relationships.
%%%%%%%%%%%%%%%%%%%%%%%%%%%%
%%% Conclusion %%%
%%%%%%%%%%%%%%%%%%%%%%%%%%%%
\section{Discussion and Conclusions}\label{sec:conclusion}
%%%%%%%%  Recap of method
We developed a physics-constrained DNN parameter estimation framework for the 0D flow battery model, which enables learning model parameters as functions of operating conditions.
In this framework, DNNs are used to approximate the map from the operating conditions to the model parameters,
and the 0D VRFB physical model combined with the DNN parameter functions is used to predict the cell voltage.
Thus, different from direct data-driven DNN methods that learn a map from the operating conditions to the voltage, the PCDNN model voltage output satisfies the given physical model.
The key idea of the proposed approach is to directly learn the parameter functions by training DNNs subject to physics model constraints with the experimental datasets that are collected from many experiments.
Therefore, the trained PCDNN model can predict both the model parameters and cell voltage under varying operating condition, including the conditions that are not part of the training dataset. 
This is different from standard model calibration approaches such as LSE, where modeling a battery under certain operating conditions requires calibrating the model for these same operating conditions.
Such model calibration approaches cannot be used to accurately predict battery performance under new conditions when the model parameters have a strong dependence on the operating conditions. 

%%%%%%%%  Achievement
% 1. show the estiamtion can be exactly if system is constant
% 2. show the improved predcition for complex combination of experiment dataset.
% 3. show the generalization for unseen data.
In this work, we demonstrated the effectiveness of the PCDNN approach by using the simulation data generated with reference operating-condition-independent parameters, as discussed in Section \ref{sec:validation_0D}. We demonstrated that the PCDNN approach can estimate such parameters more accurately than the LSE approach.
% \todo{I do not think we adequately explained why PCDNN does better than LSE for this synthetic test problem. QH: Table 4 shows that PCDNN can find the right parameters while LST doesn't given the same normalization. AT: We demosntrated that PCDNN is more accurate but we have not explained why.}
%, suggesting its enhanced capacity of mitigating ill-posedness issues in cases of multiple parameter estimation.
We also demonstrated that the PCDNN method produces equally accurate results for problems with three and four unknown parameters, while the accuracy of the LSE method decreases with an increasing number of unknown parameters. 
%yields almost exact parameter estimation in the case with 3 identifiable parameters ($S, k_n, \sigma_e$).
%It shows that the proposed approach can accurately reproduce constant functions if the parameters corresponding to the underlying physical model are independent of the operating conditions.
In Section \ref{sec:res_exp}, we tested the PCDNN approach for an experimental dataset consisting of voltage measurements as functions of time and operating conditions from 12 different experiments.
The results show that the PCDNN approach significantly improves the voltage prediction compared to the predictions of the 0D model with parameters reported in the literature \cite{Chen2021} and estimated from the LSE approach. 
The PCDNN method produced more accurate predictions for both the experiments that were used for parameter estimation and the \emph{unseen} experiments, i.e., those experiments that were not used in the training of the PCDNN model. % By automatically adjusting the parameter outputs with respect to the given operating conditions, we have also shown that the trained PCDNN model can be used to predict cell voltage for different experiments not included in the training datasets.
These results demonstrate the enhanced abilities of parameter estimation and predictive generalization offered by the PCDNN approach, which provides a flexible framework to leverage the information of physical models and experimental data. 
It also serves as a surrogate model that allows efficient parameter estimation and avoids repeated, time-consuming calibration procedures.

%%%%%%%%  Limitation & Future
In this work, we used the PCDNN approach to estimate the functional forms of parameters in a simple 0D model that has a limited ability to capture the tails of the charge and discharge curves, noticeable discrepancies in cell voltage predictions were observed at the extreme SOC values in some of the experiments. Considering additional physics in the 0D model, such as the concentration overpotential, can further improve the cell voltage prediction of the ``PCDNN-0D'' model. 
\textcolor{red}{We should note that the PCDNN approach can be also used to estimate the functional form of parameters in   higher-dimensional physics-based models of flow batteries that would improve these models' accuracy with respect to standard LSE parametarization; however, this would also require numerically solving the PDEs or approximating the solution of PDEs with a DNN as in the standard PINN method.}

Finally, we note that introducing more physics (e.g., concentration overpotential and cross-over) and considering more changing model parameters, such as membrane properties, electronic conductivity, and diffusion coefficients in the electrolyte) in the 0D model could enable the PCDNN method to predict more complex aspects of battery performance, such as degradation. 

% membrane diffusiion, migration, and convection will affect the degration.
%including the PINN method can account for   In the future, the proposed approach can consider the effect of degradation as the charge and discharge voltage responses are altered during operating, indicating the .
%The PINN approach could be a promising tool to estimate these parameters and help to advance the design of long life-low cost redox flow battery. 

%%% ---------------------------------
% Acknowledgement
%%% ---------------------------------
\section{Acknowledgments}
The authors thank Litao Yan, Yunxiang Chen, and Jie Bao for helpful discussions. The work was supported by the Energy Storage Materials Initiative (ESMI), which is a Laboratory Directed Research and Development Project at Pacific Northwest National Laboratory. Pacific Northwest National Laboratory is operated by Battelle for the U.S. Department of Energy under Contract DE-AC05-76RL01830.
%%%%%%%%%%%%%%%%%%%%%%%%%%%%
%%% appendix %%%
%%%%%%%%%%%%%%%%%%%%%%%%%%%%
\appendix

%%% ---------------------------------
% 0D models
%%% ---------------------------------
\section{Analytical solution of species concentrations}\label{sec:append-concentration}
Following the 0D VRFB model proposed in \cite{Shah2011}, the species considered in the reaction kinetics are $\mathcal{S} = \{\text{V(II)}, \text{V(III)}, \text{V(IV)}, \text{V(V)}, \text{H}^+, \text{H}_2 \text{O}\}$. 
Let $c_i^{res}$ and $c_i$ be the concentrations of species $i$ ($ i \in \mathcal{S}$) in the reservoir and the electrode, respectively. Note that both $c_i^{res}$ and $c_i$ are space-independent functions because of the assumption of uniform distribution in the 0D model.
% \todo{QH: I deleted the following commented sentence.}
% Thus, the recirculation of the electrolytes through reservoirs, changes the concentrations of all species in the electrodes together with the electrochemical reaction.

With a uniform flow rate at the inlet, the net change per unit time ($\text{mol/s}$) of species $i$ in the electrode due to recirculation is $\epsilon A_{in} u ( c_i^{res} - c_i)$, where $\epsilon$ is the porosity of the electrode, $u$ is the electrolyte flow velocity (in $\text{m/s}$), and $A_{in} = b_e w_e$ is the inlet area of the electrode, in which $b_e$ and $w_e$ are the breadth and width, respectively.
% \todo{QH: I deleted the following commented sentence.}
% (see Fig. \ref{fig:sche_model} and the default values given in Table \ref{table:ex1_model}).
Here, we define $\tilde{u} = \epsilon u =  \omega /A_{in}$ as the average velocity in the porous medium, where $\omega$ is the inlet volume flow rate. Therefore, the mass conservation of species $i$ in the reservoirs is:
\begin{equation}\label{eq:mass_res}
V_r \frac{d c_i^{res}}{dt} = - A_{in} \tilde{u} ( c_i^{res} - c_i)
%= - \frac{\epsilon \delta u}{h_e} ( c_i^{res} - c_i), \quad i \in \mathcal{S}
\end{equation}
where $V_r$ is the volume of the reservoir.

In the electrodes, we consider both the electrochemical reaction and the recirculation in the mass balance equations for the species ($i = \text{V(II), V(III), V(IV), V(V)}$), expressed as:
\begin{subequations}\label{eq:mass_ele}
	\begin{align}
		& \epsilon V_e \frac{d c_i}{dt} = A_{in} \tilde{u} ( c_i^{res} - c_i) - \frac{I}{F}, \quad i = \text{V(II), V(V)}, \\
		& \epsilon V_e \frac{d c_i}{dt} = A_{in} \tilde{u} ( c_i^{res} - c_i) + \frac{I}{F}, \quad i = \text{V(III), V(IV)},
	\end{align}
\end{subequations}
%\begin{equation}\label{eq:kine_v2}
%\epsilon V_e \frac{d c_i}{dt} = A_{in} \tilde{u} ( c_i^{res} - c_i) - \frac{I}{F}, \quad i = V(II), V(V)
%\end{equation}
%\begin{equation}\label{eq:kine_v3}
%\epsilon V_e \frac{d c_i}{dt} = \epsilon A_{in} u ( c_i^{res} - c_i) + A_s \frac{j_{app}}{F}, \quad i = \text{V(III), V(IV)}
%\end{equation}
where $I$ is the applied current, $V_e = b_e w_e h_e$ is the volume of the electrode, and $h_e$ is the length of the electrode. Eliminating the recirculation terms in Eqs. (\ref{eq:mass_res}) and (\ref{eq:mass_ele}) and integrating time using the initial conditions, the solutions for electrode concentrations can be obtained \cite{Shah2011}:
\begin{subequations}\label{eq:soln_ele}
	\begin{align}
	& c_i = c_i^0 - \frac{I}{V_e F \epsilon \tilde{ \epsilon} } \left(\frac{\epsilon \delta + e^{-\tilde{ \epsilon} t}}{1+ \epsilon \delta} - 1 - \frac{\epsilon \delta}{\tau} t \right), \quad i = \text{V(II), V(V)}, \\
	& c_i = c_i^0 + \frac{I}{V_e F \epsilon \tilde{ \epsilon}} \left(\frac{\epsilon \delta + e^{-\tilde{ \epsilon} t}}{1+ \epsilon \delta} - 1 - \frac{\epsilon \delta}{\tau} t \right), \quad i = \text{V(III), V(IV)},
	\end{align}
\end{subequations}
where $c_i^0$ are the initial concentrations for both $c_i^{res}$ and $c_i$, $\delta = V_e/V_r$ is the ratio of the two volumes, $\tau = h_e/u$ and $\tilde{\epsilon} = (\epsilon \delta + 1)/\tau$.
% With the concentration of vanadium specified, 
Then, we can define the state of charge (SOC) \cite{You2009a} as
% An important factor of the charge or discharge process is State of Charge (SOC) which introduces the amount of electrochemical energy stored in the tanks, and is described as follows:
\begin{equation}\label{eq:soc}
\text{SOC}(t) = \frac{c_{\text{V(II)}}(t)}{\bar{c}_{V_n}} = 1 -  \frac{c_{\text{V(III)}}(t)}{\bar{c}_{V_n}}
\end{equation}
where $\bar{c}_{V_n}$ is the
% initial vanadium concentration
total vanadium concentration
of the negative half-cell. 
% \todo{QH: I deleted the following commented sentence because it is redundant.}
% If $c_{\text{V(III)}}^0 = \bar{c}_{V_n}$ and 
Substituting the analytical solution for $c_{\text{V(II)}}$ from Eq. (\ref{eq:soln_ele}a) into Eq. (\ref{eq:soc}), SOC can be rewritten as
\begin{equation}\label{eq:soc_t}
\text{SOC}(t) = \text{SOC}^0 - \frac{I}{\bar{c}_{V_n} V_e F \epsilon \tilde{ \epsilon} } \left(\frac{\epsilon \delta + e^{-\tilde{ \epsilon} t}}{1+ \epsilon \delta} - 1 - \frac{\epsilon \delta}{\tau} t \right)
\end{equation}
where $\text{SOC}^0 = c_{\text{V(II)}}^0 / \bar{c}_{V_n}$ is the initial SOC.
% The SOC estimated by Eq. (\ref{eq:soc})
% % and (\ref{eq:soc_t})
% is used to calculate the species concentrations in Eq. \eqref{eq:SOC_pinn}.

A similar procedure is adopted for the concentrations of water and protons, i.e., $c_{\text{H}_2 \text{O}}$ and $c_{\text{H}^+}$.
% Considering the recirculation,  electrochemical reaction, and osmotic drag effect, 
The mass conservation of water is
\begin{subequations}\label{eq:mass_water}
	\begin{align} 
%	c_{\text{H}_2 \text{O}}
	& - \text{ve electrode}: \epsilon V_e \frac{d c_{\text{H}_2 \text{O}}}{dt} = A_{in} \tilde{u} ( c_{\text{H}_2 \text{O}}^{res} - c_{\text{H}_2 \text{O}}) + \frac{ n_d I}{F},  \\
	& + \text{ve electrode}: \epsilon V_e \frac{d c_{\text{H}_2 \text{O}}}{dt} = A_{in} \tilde{u}  ( c_{\text{H}_2 \text{O}}^{res} - c_{\text{H}_2 \text{O}}) -  \frac{(1+ n_d) I}{F}.
	\end{align}
\end{subequations}
Herein, the molar flux of water through the membrane from the positive to negative electrode during charging is approximated by $n_d j_{app}/F$, where $n_d$ is the drag coefficient, $j_{app} = I/A_e$ is the nominal current density, and $A_e = h_e w_e$ is the electrode area. By using the mass balance equation in (\ref{eq:mass_res}) with $i = \text{H}_2 \text{O}$ to eliminate the recirculation terms, the solution to Eq. (\ref{eq:mass_water}) reads
\begin{subequations}\label{eq:soln_water}
	\begin{align}
	& - \text{ve electrode}:  c_{\text{H}_2 \text{O}} = c_{\text{H}_2 \text{O}}^0 
	- \frac{I n_d}{V_e F \epsilon \tilde{ \epsilon} }
	\left(\frac{\epsilon \delta + e^{-\tilde{ \epsilon} t}}{1+ \epsilon \delta} - 1 - \frac{\epsilon \delta}{\tau} t \right),\\
	& +\text{ve electrode}: c_{\text{H}_2 \text{O}} = c_{\text{H}_2 \text{O}}^0 
	+ \frac{I}{V_e F \epsilon \tilde{ \epsilon}} (1+n_d)
	\left(\frac{\epsilon \delta + e^{-\tilde{ \epsilon} t}}{1+ \epsilon \delta} - 1 - \frac{\epsilon \delta}{\tau} t \right).
	\end{align}
\end{subequations}
Using the SOC in Eq. (\ref{eq:soc_t}), the water concentration in the positive electrode $c_{\text{H}_2 \text{O}_p}$ is expressed as: $c_{\text{H}_2 \text{O}_p} = c_{\text{H}_2 \text{O}_p}^0 - (1 + n_d) \bar{c}_{V_n} \times (\text{SOC} - \text{SOC}^0)$. The mass balances for the protons are:
\begin{subequations}\label{eq:mass_proton}
	\begin{align} 
	%	\text{H}+
	& - \text{ve electrode}: \epsilon V_e \frac{d c_{\text{H}^+}}{dt} = A_{in} \tilde{u} ( c_{\text{H}^+}^{res} - c_{\text{H}^+}) + \frac{I}{F},  \\
	& + \text{ve electrode}: \epsilon V_e \frac{d c_{\text{H}^+}}{dt} = A_{in} \tilde{u}  ( c_{\text{H}^+}^{res} - c_{\text{H}^+}) +  \frac{I}{F}.
	\end{align}
\end{subequations}
For simplicity, in the above equation (\ref{eq:mass_proton}) we assume a rapid transport of protons across the fully saturated membrane such that the protons are evenly distributed in both the positive and negative electrodes, which is different from the original formulation in \cite{Shah2011}. 
Thus, the proton concentrations in both electrodes are:
\begin{equation}\label{eq:soln_proton}
	 c_{\text{H}^+} = c_{\text{H}^{+}}^0
	 - \frac{I}{V_e F \epsilon \tilde{ \epsilon} }
	 \left(\frac{\epsilon \delta + e^{-\tilde{ \epsilon} t}}{1+ \epsilon \delta} - 1 - \frac{\epsilon \delta}{\tau} t \right),
\end{equation}
or expressed as $c_{\text{H}^+} = c_{\text{H}^{+}}^0 + \bar{c}_{V_n} \times (\text{SOC} - \text{SOC}^0)$.
%- \text{ve electrode} \text{and}  + \text{ve electrode}:

For the experiments considered in this study, we set the initial concentrations $c_{\text{V(III)}}^0 = c_{\text{V(IV)}}^0 = c_{V}^0$ and $c_{V}^0 = \bar{c}_{V_n}$. Thus, we have $\text{SOC}^0 = 0$.
Then, the SOC estimated by Eqs. (\ref{eq:soc}) and (\ref{eq:soc_t}) is used to calculate the concentration of each species, as shown in Eq. \eqref{eq:SOC_pinn}.

\section{Deep neural network}\label{sec:appendix-DNN-appr}
% \todo{QH:Delete the figure B.9 due to the limit of figures.}
% \begin{figure}[ht!]
% 	\centering
% 	\includegraphics[angle=0,width=4in]{manuscript-PINN-VFB-I-v1/figures/Scheme/DNN.pdf}
% 	\caption{Schematic representation of a feed-forward deep neural network.}
% 	\label{fig:DNN}
% \end{figure}

In the proposed approach, we use a fully connected feed-forward network architecture known as multilayer perceptrons, where the basic computing units (neurons) are stacked in layers.
% as shown in Fig. \ref{fig:DNN}. 
Generally, the DNN approximation $\hat{\vec{u}}(\vec{x};\theta)$ of a function $\vec{u}(\vec{x})$ is given as:
\begin{equation}
\vec{u}(\vec{x}) \approx \hat{\vec{u}}(\vec{x};\theta)  = \vec{y}_{n_l+1} (\vec{y}_{n_l}(...(\vec{y}_2(\vec{x}))),
\end{equation}
where $\hat{(\cdot)}$ denotes the DNN approximation, and
\begin{equation}
\begin{split}
\vec{y}_2(\vec{x}) &= \sigma(\vec{W}_1 \vec{x} + \vec{b}_1) \\
\vec{y}_3(\vec{y}_2) &= \sigma(\vec{W}_2 \vec{y}_2 + \vec{b}_2)\\
... \\
\vec{y}_{n_l} (\vec{y}_{n_l-1}) &= \sigma(\vec{W}_{n_l-1} \vec{y}_{n_l-1} + \vec{b}_{n_l-1})\\
\vec{y}_{n_l+1} (\vec{y}_{n_l}) &= \vec{W}_{n_l} \vec{y}_{n_l} + \vec{b}_{n_l}.
\end{split}
\end{equation}
The first layer is called the input layer, and the last layer is the output layer, while all the intermediate layers are known as hidden layers.
Here, $n_l$ denotes the number of hidden layers, $\sigma$ is the predefined activation function, $\vec{x} \in \mathbb{R}^d$ denotes the input ($d$ is the number of spatial dimensions), $\vec{y}_{n_l+1} $ is the output vector, and $\theta$ denotes all the parameters (weights and biases) in the DNN approximation of $\vec{u}$:
\begin{equation}
\theta = \{\vec{W}_1, \vec{W}_2, ..., \vec{W}_{n_l}, \vec{b}_1, \vec{b}_2, ..., \vec{b}_{n_l}\}.
\end{equation}
%\textcolor{red}{In the "data-driven" approach, $\theta$ is directly estimated from the measurements of $\vec{u}$ by minimizing the loss function  
%	$L(\theta) = \sum_{\vec{x} \in \mathcal{T}_u} (\hat{\vec{u}} (\vec{x}; \theta) - \vec{u}^*(\vec{x}))^2$:}
%\begin{equation}\label{loss_data}
%\theta = \text{arg}\min \limits_\theta  \sum_{\vec{x} \in \mathcal{T}_u} (\hat{\vec{u}} (\vec{x}; \theta) - \vec{u}^*(\vec{x}))^2,
%\end{equation}
%where  $\mathcal{T}_u = \{\vec{x}_1,\vec{x}_2,...,\vec{x}_{|\mathcal{T}_u|} \}\subset \Omega$ denotes a set of measurement locations, $\Omega \subset \mathbb{R}^d$ is the domain of the function $\vec{u}$, and ${\vec{u}^*(\vec{x}), \vec{x} \in \mathcal{T}_u}$ are the measured values of $\vec{u}$ at these locations.

%\section{Nomenclature}
%\nomenclature{$a$}{The number of angels per unit area}
%\nomenclature{$N$}{The number of angels per needle point}
%\nomenclature{$A$}{The area of the needle point}
%\printnomenclature[0.6in]

%\nomenclature{$a$}{The number of angels per unit area}
%\nomenclature{$N$}{The number of angels per needle point}
%\nomenclature{$A$}{The area of the needle point}
%
%\printnomenclature[0.6in]
%\begin{itemize}
%	\item [] $\epsilon$ \quad \quad Electrode porosity
%	\item [] $\eta^{act}$ \quad \quad Over-potential
%\end{itemize}

% ===============================================================================
%%%%%%%%%%%%%%%%%%%%%%%%%%%%%%%%%%%%%%%%%%%%%%%%%%%%
%%%     End of body of article     %%%
%%%%%%%%%%%%%%%%%%%%%%%%%%%%%%%%%%%%%%%%%%%%%%%%%%%%
%\section*{Acknowledgements}

\bibliography{references.bib,mybib_PINN.bib}
% ===============================================================================

\end{document}